\documentclass[fleqn,usenatbib]{mnras}
\setcitestyle{notesep={ }} 

\usepackage{newtxtext,newtxmath}

\usepackage[T1]{fontenc}

\DeclareRobustCommand{\VAN}[3]{#2}
\let\VANthebibliography\thebibliography
\def\thebibliography{\DeclareRobustCommand{\VAN}[3]{##3}\VANthebibliography}
\newcommand{\dotdeg}{\rlap{.}^\circ}

\usepackage{graphicx}	
\usepackage{amsmath}	
\usepackage{multirow}
\usepackage[round-mode=places,
            separate-uncertainty=true,
            table-align-uncertainty=true]{siunitx}
\usepackage[table]{xcolor}

\definecolor{ProjRed}{RGB}{255,0,0}
\definecolor{ProjOrange}{RGB}{255,179,96}
\definecolor{ProjGreen}{RGB}{128, 255, 180}
\definecolor{ProjGreenDarker}{RGB}{102, 204, 143}
\definecolor{ProjBlue}{RGB}{0, 181, 235}
\definecolor{ProjPurple}{RGB}{128, 0, 255}

\title[The Cosmic Dipole in Quaia]{The Cosmic Dipole in the Quaia Sample of Quasars: A Bayesian Analysis}

\author[Mittal, Oayda $\&$ Lewis]{
Vasudev Mittal,$^{1,2*}$\thanks{E-mail: vasudeviiser@gmail.com}
Oliver T. Oayda,$^{2*\thanks{E-mail: ooay3125@uni.sydney.edu.au}}$
and Geraint F. Lewis${^2}$
\\
$^{*}$Joint first author\\
$^{1}$Department of Physical Sciences, IISER Mohali, Knowledge City, Sector 81, SAS Nagar, Manauli PO 140306, Punjab, India\\
$^{2}$Sydney Institute for Astronomy, School of Physics A28, The University of Sydney, NSW 2006, Australia\\
}

\date{Accepted 2023 November 24. Received 2023 November 15; in original form 2023 October 17}

\pubyear{2023}

\begin{document}
\label{firstpage}
\pagerange{\pageref{firstpage}--\pageref{lastpage}}
\maketitle

\begin{abstract}
    We present a Bayesian analysis of the Quaia sample of 1.3 million quasars as a test of the cosmological principle.
    This principle postulates that the universe is homogeneous and isotropic on sufficiently large scales, forming the basis of prevailing cosmological models.
    However, recent analyses of quasar samples have found a matter dipole inconsistent with the inferred kinematic dipole of the Cosmic Microwave Background (CMB), representing a tension with the expectations of the cosmological principle.
    Here, we explore various hypotheses for the distribution of quasars in Quaia, finding that the sample is influenced by selection effects with significant contamination near the galactic plane.
    After excising these regions, we find significant evidence that the Quaia quasar dipole is consistent with the CMB dipole, both in terms of the expected amplitude and direction.
    This result is in conflict with recent analyses, lending support to the cosmological principle and the interpretation that the observed dipole is due to our local departure from the Hubble flow.
\end{abstract}

\begin{keywords}
    quasars: general -- cosmology: observations -- cosmology: theory -- large-scale structure of Universe
\end{keywords}



\section{Introduction}%
    A critical assumption in the contemporary cosmological framework is that the universe is homogeneous and isotropic at the largest scales \citep{harrison2000}.
    This is the cosmological principle, and it is for example taken as a starting point by the Friedmann-Lema\^{i}tre-Robertson-Walker (FLRW) metric of spacetime and the Friedmann equations describing cosmic evolution.
    Homogeneity and isotropy were initially raised to the level of an \textit{a priori} principle by \citet{milne1935} -- but the question as to whether there is an \textit{a posteriori} basis remains.
    If such a basis cannot be found, then we must critically re-examine the support for prevailing cosmological models.
    
    The cosmological principle tacitly assumes the existence of a set of fundamental observers which reside in the `cosmic rest frame' where the universe is maximally isotropic.
    This is supported by the fact that the `cosmic microwave background' (CMB) is remarkably smooth with temperature anisotropies of order $\Delta T / T \approx 10^{-5}$. 
    However, imprinted on these underlying small-scale fluctuations is a dipole anisotropy of order $\Delta T / T \approx 10^{-3}$. 
    This is conventionally explained by the Earth's peculiar motion through the universe with a speed of $369.82\pm0.11\,\text{km}\,\text{s}^{-1}$ towards $(l,b) = (264\dotdeg021, 48\dotdeg253)$ in galactic coordinates \citep{planck2020}, which we denote as $\mathbf{v}_\text{CMB}$ for future reference.
    If this explanation (the kinematic interpretation of the CMB) is correct, then other cosmological probes using all-sky surveys should show a similar anisotropy.
    Critically, distributions of matter at sufficiently large distances -- namely where local clustering effects are negligible -- should exhibit a dipole anisotropy, which we call the `cosmic dipole' or the `matter dipole'.
    If the cosmological principle is an accurate description of the universe, then the peculiar velocity inferred from this dipole should correspond with $\mathbf{v}_\text{CMB}$.
    
    This matter anisotropy is observed, but there is no clear consensus on whether it is consistent with the cosmological principle or not.
    However, the general trend is that the matter dipole studies -- specifically with radio galaxies and quasars -- find a dipole that aligns with the CMB dipole in direction, but is larger in magnitude \citep{Peebles_2022, aluri2023}.
    This `dipole anisotropy problem' thus represents an outstanding problem amongst cosmological probes.
    Insofar that a consensus on this issue has not been reached, independent studies of matter dipoles with new catalogues of sources are key in further understanding the nature of this anomaly; for example, does it represents a shortcoming of our scientific understanding or an as of yet unresolved systematic issue?

    With this in mind, in this work we present an analysis of the recently-released Quaia quasar catalogue \citep{storeyfisher2023quaia}.
    At the highest magnitude limit, this catalogue contains $1\,295\,502$ sources.
    We examine the anisotropy in angular distribution of these quasars over the sky, applying a Bayesian framework to compare the inferred dipole to that of the CMB.
    The structure of this paper is as follows.
    In Section~\ref{sec:background}, we present the background theory and an overview of the instant state of the literature, including current observations of the cosmic dipole and an assessment of their consistency with the cosmological principle.
    The data under consideration in this study -- the Quaia quasar catalogue -- is presented in Section~\ref{sec:quaia-catalogue}, and our approach to analysing the sample is examined in Section~\ref{sec:approach}.
    The results are presented in Section~\ref{sec:results}.
    We discuss our results and present our conclusions in Section~\ref{sec:discussion}.

\section{Background: Number count dipole}\label{sec:background}
    The cosmological principle's key assumption of homogeneity and isotropy can, and has been, tested.
    One critical family of tests involves probing the distribution of matter in the Earth's frame of reference; these are the `matter dipole' studies.
    If we assume the principle to be an accurate description of the universe, then the CMB's temperature dipole is interpreted to arise solely from the Earth's peculiar motion.
    Moreover, the dipole-removed frame is the frame of `cosmic rest' where the universe is perceived as maximally isotropic and homogeneous.
    Insofar that the Earth's peculiar velocity imprints a Doppler shift on observed sources like radio galaxies, we should be able to recover the magnitude and direction of this motion from the dipole in matter distributions over the sky.
    Framed in this way, measuring the consistency between the CMB-inferred and matter-inferred velocities is the linchpin of the matter dipole studies.

    To see this, consider an observer moving with velocity $v \ll c$ with respect to distant sources which are isotropic and homogeneous in their own rest frame.
    As suggested in \cite{ellis1984}, if within the observer's passband the sources have a spectral energy distribution with a power law dependence on frequency described by $S \propto \nu ^{-\alpha}$, and the apparent flux density has a cumulative power law distribution $N(>S) \propto S^{-x}$, then Doppler boosting and relativistic aberration will induce a dipole anisotropy in the distribution of sources in the observer's frame.
    The isotropic frame of reference will be boosted by an amplitude
    \begin{equation}
        {\bf \mathcal{D}} = [2 + x(1+\alpha)]\frac{v}{c}. \label{eq:dipole-magnitude}
    \end{equation}
    This is the famous `kinematic dipole', and \citet{ellis1984} made the rough estimate that a minimum of $O(10^{5})$ sources would be needed to discern this dipole.
    The implicit assumption here is that the observer should survey the sky until a flux density above which there is no directional bias in the completeness of the survey.
    Additionally, $x$ and $\alpha$ are assumed to not be redshift-dependent, although there has been some suggestion that this simplification should be revisited \citep[see e.g.][]{dalang2022}.
    Further, local inhomogeneities introduce a clustering dipole, so for a genuine measurement of the cosmic dipole a significant fraction of the sources need to be at high redshifts \citep[$z\approx 1$;][]{tiwari2016}.
    From equation~\eqref{eq:dipole-magnitude}, the net dipole anisotropy for a patch of sky in the direction $\bf \hat{n}$ will be
    \begin{equation}
        \frac{\Delta N}{N} = {\bf D}\cdot{\bf \hat{n}} = [2 + x(1+\alpha)]\frac{\bf v}{c}\cdot{\bf \hat{n}}. \label{eq:E&B-dipole}
    \end{equation}
    Various all-sky surveys of radio sources have been used to trace out this dipole over the sky, and thus probe the cosmological principle.
    We note that \citet{aluri2023} deals with the genealogy of these tests in greater detail, but none the less we recount some of the salient results here.
    
    \citet{blake2002} initially found support for a kinematic dipole aligned with the CMB and possessing the expected amplitude.
    However, the immediate state of the literature is equivocal as to whether or not the matter dipole is consistent with the CMB dipole.
    Many studies \citep[see e.g.][]{singal2011, colin2017, bengaly2018, singal2019, siewert2021, singal2023, wagenveld2023} have reported dipole amplitudes that are in excess of the CMB expectation, while the inferred dipole directions generally align with the CMB dipole (although notably \citet{darling2022} and \citet{cheng2023} find consistency with the CMB dipole for their chosen radio catalogues).
    We point out that in the foregoing works and amongst others, authors discussed the appropriate choice of dipole estimator at length, including whether or not certain estimators incur a bias that must be accounted for.
    To our knowledge, tests instead formulated in the language of Bayesian statistics have been used less extensively, which we discuss below.
    
    Turning away from the radio galaxy studies, \cite{secrest2021} showcased that the \citet{ellis1984} method can be used to study the matter dipole in quasar samples.
    This study, taken together with the joint radio galaxy and quasar analysis in \cite{secrest2022}, is perhaps one of the more significant challenges to the cosmological principle.
    Therein, the authors studied the dipole in the distribution of quasars from CatWISE2020 \citep{marocco2021} using a least squares estimator, finding that the amplitude was at least twice as large as expected (at a 4.9$\sigma$ level of statistical significance).
    A similar conclusion with the same sample was reached in \citet{kothari2022}.
    Separately, \citet{singal2021} used a sample of 0.28 million quasars and also found a dipole magnitude in excess of the CMB expectation, although the sample size there was about 5 times smaller than that of \citet{secrest2021}.

    As we touched on earlier, these analyses used frequentist statistics, and the results are sensitive to the estimator chosen.
    However, a Bayesian analysis of CatWISE2020 was performed by \cite{dam2023}, in which \citet{secrest2021}'s result of an anomalously large dipole was confirmed at a statistical significance of $5.7\sigma$.
    Taken together, these results lend evidence to the proposition that the quasar dipole is in tension with the kinematic dipole inferred from the CMB.

    On the basis of the foregoing, the literature interrogating the matter dipole is by no means unanimous.
    That being said, these works do not represent an exhaustive survey of what is possible; a suite of other probes have been formulated, many of which are accounted for in \citet{aluri2023}.
    Some of these include tests with Type Ia SNe \citep[see e.g.][]{horstmann2022, singal2022, sorrenti2022}, analyses of bulk flows \citep[see e.g.][]{watkins2023} and direct probes of the FLRW metric with tests of spatial curvature \citep[see e.g.][]{zhou2020}.
    Recently, \citet{oayda2023} proposed a novel test involving a dipole in time dilation, as sources with intrinsic time-scales are time dilated along the direction of the Earth's motion.
    
    Returning to quasars, if there is an outstanding tension between the dipole inferred from quasars and that expected from the CMB, then closer scrutiny is warranted.
    Since the cosmological principle is a foundational assumption in the prevailing cosmological paradigm \citep{harrison2000}, a challenge to it cannot be easily overlooked.
    In this work, we present another analysis of the dipole in quasar distributions.
    We tested the recently-released Quaia catalogue \citep{storeyfisher2023quaia}, employing Bayesian inference to understand which model is best supported by the sample and whether the inferred dipole is consistent with that of the CMB.

\section{Quaia catalogue}
    \label{sec:quaia-catalogue}
    The Quaia catalogue \citep{storeyfisher2023quaia} is principally taken from quasars observed by the \textit{Gaia} satellite \citep{gaia2016}, which were released in \textit{Gaia} DR3 \citep{gaia2023b,gaia2023a}.
    The full sample of DR3 quasar candidates totals to $6\,649\,162$ sources, which was the starting point for the authors.
    
    In constructing their catalogue, they first imposed that all \textit{Gaia} quasars have a measurement of photometric magnitude in the $G$, $BP$ and $RP$ bands.
    Additionally, the authors cross-matched each of the quasar candidates with those from the \textit{Wide-field Infrared Survey Explorer} \citep[\textit{WISE};][]{wright2020}, using the unWISE reprocessing to also provide photometric information in the $W1$ and $W2$ infrared bands.
    To decontaminate their sample, the authors imposed proper motion cuts, since quasars are anticipated to be sources well within the background, and a number of colour magnitude cuts.
    They finally applied a $G < 20.5$ magnitude cut, the result of which constitutes their primary catalogue: the `Quaia high' catalogue.
    Another cut of $G < 20.0$ created the `Quaia low' catalogue, since the authors noted that deeper magnitudes sacrificed purity and measurement precision.
    
    One other issue is outstanding: selection effects.
    To mitigate these, the authors created a selection function to account for how some sources are preferentially observed at different locations on the sky due to dust extinction, stellar density and the peculiarities of \textit{Gaia}'s scanning pattern.
    This information is encoded in four maps: a dust extinction map; a stellar distribution map; a separate Large Magellanic Cloud and Small Magellanic Cloud stellar map; and, a map encoding \textit{Gaia}'s scanning law and source crowding.
    This data is passed to a Gaussian process, producing a probability map: the selection function.
    The selection function describes how likely it is for sources to be included in the final catalogue depending on where they are on the sky.
    In other words, regions which are less dense on the basis of systematics like dust extinction will be associated with a lower probability, and regions which do not suffer from these effects have a probability closer to 1.
    
\begin{figure*}
        \centering
        \includegraphics[height=5.5cm]{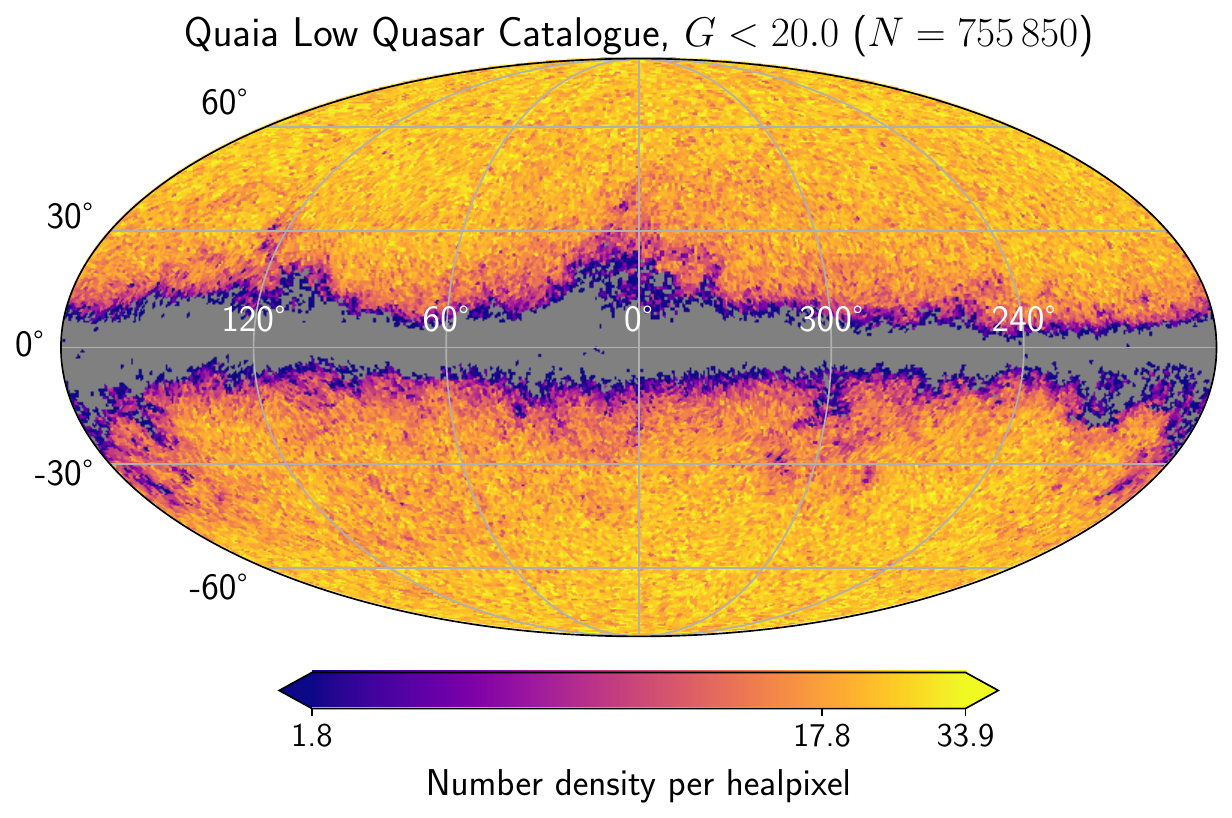}
        \hfill
        \includegraphics[height=5.5cm]{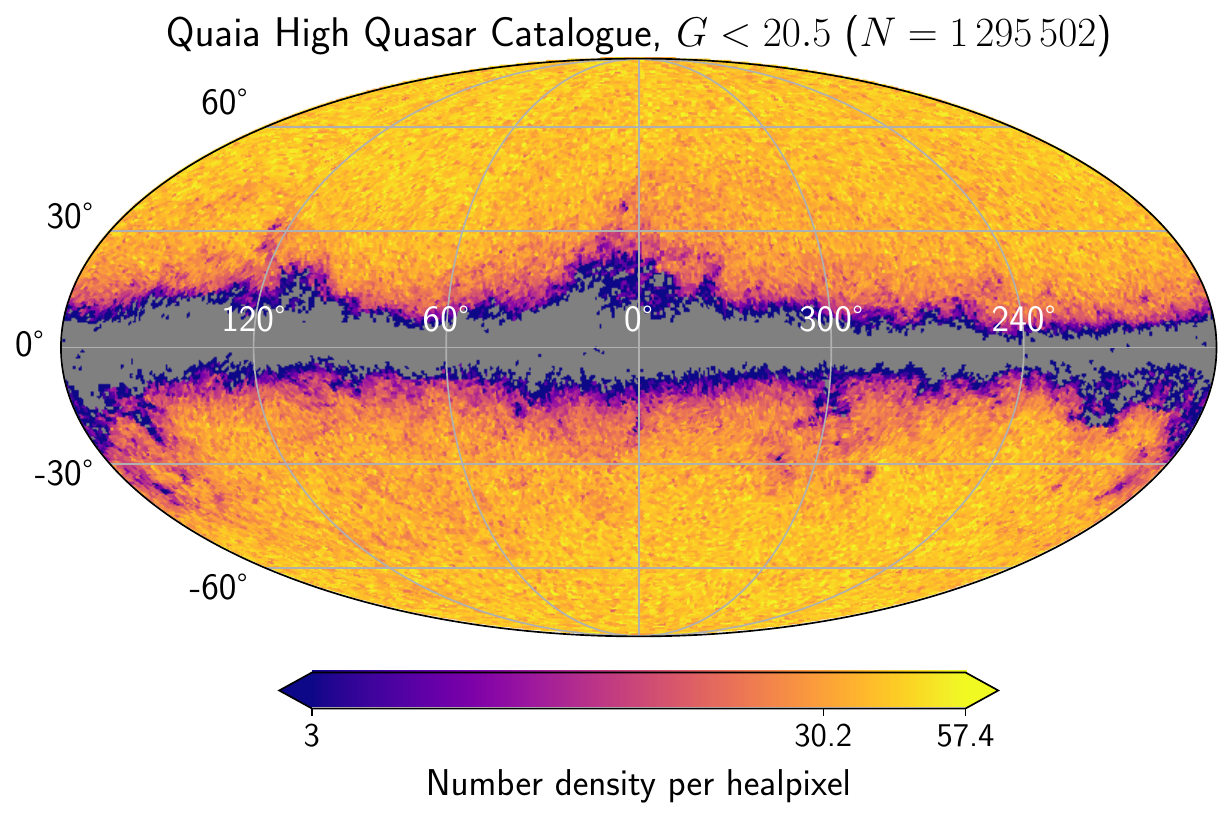}
        \includegraphics[height=5.5cm]{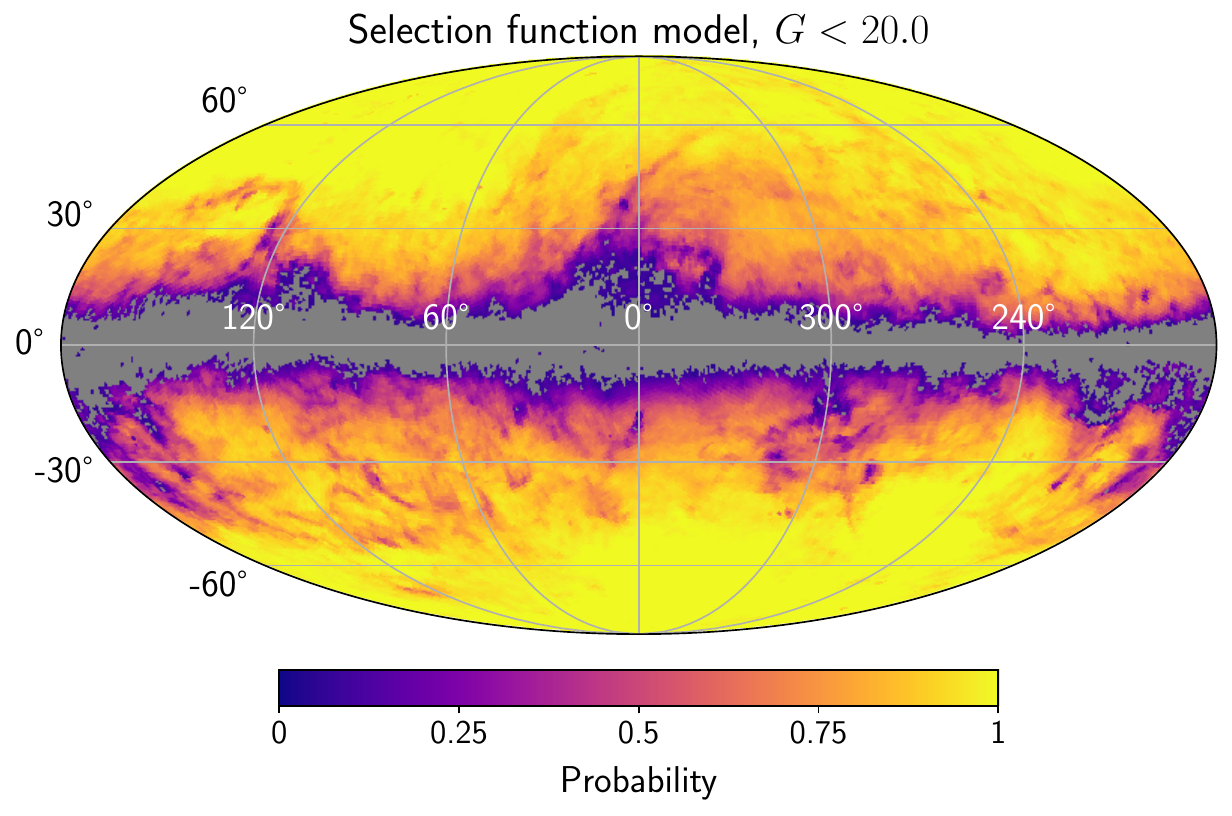}
        \hfill
        \includegraphics[height=5.5cm]{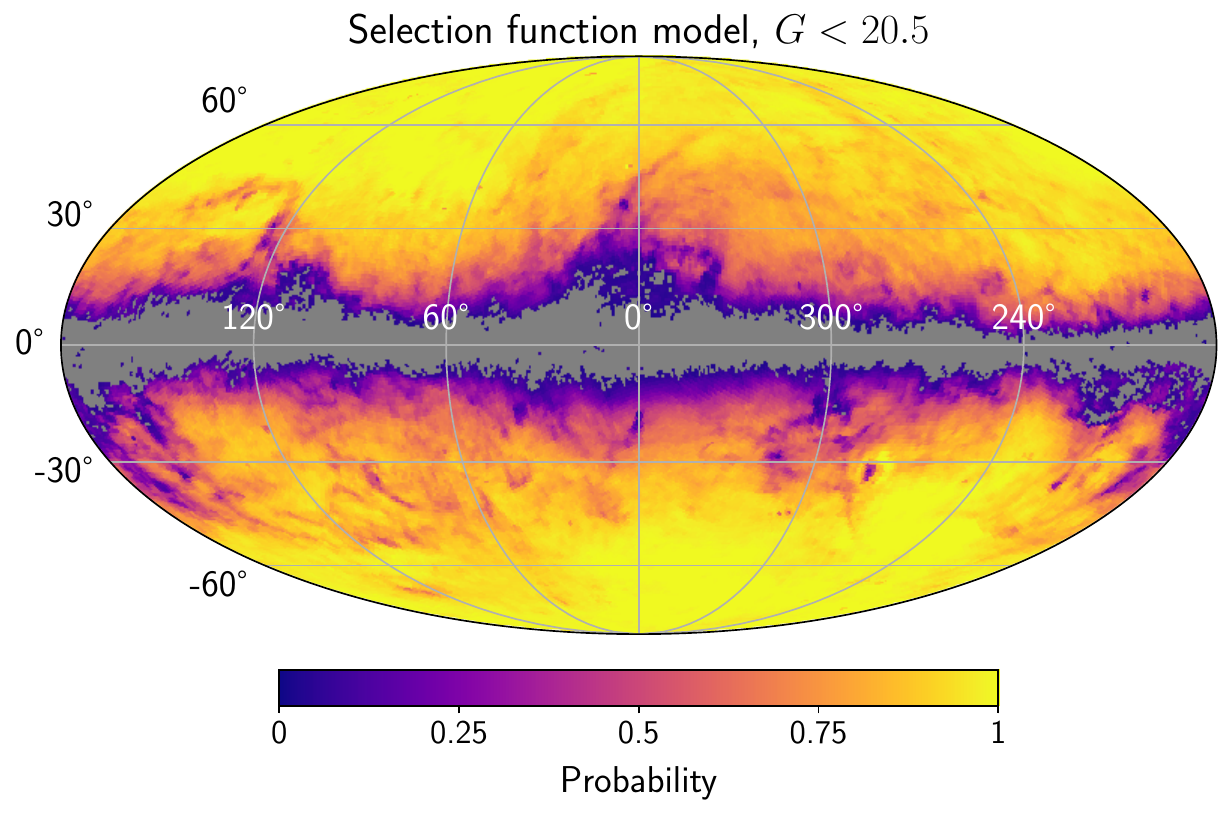}
        \includegraphics[height=5.5cm]{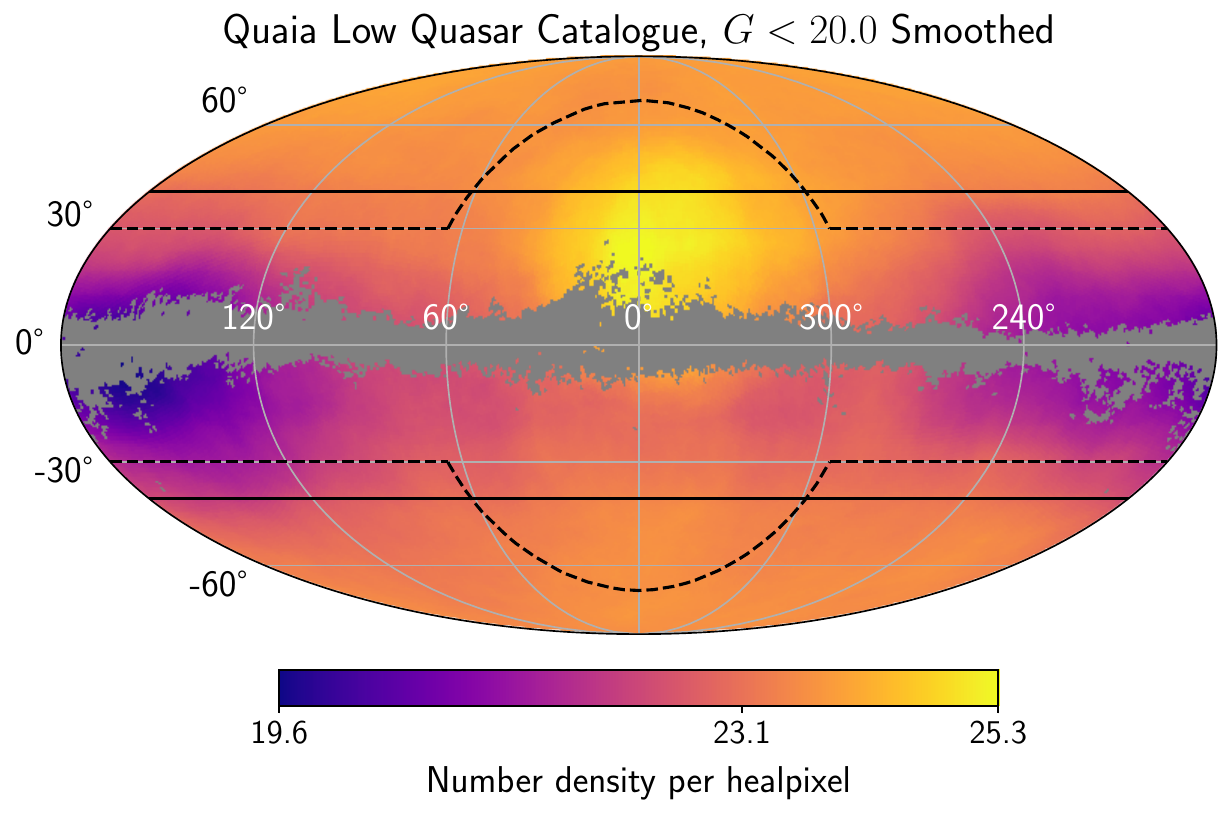}
        \hfill
        \includegraphics[height=5.5cm]{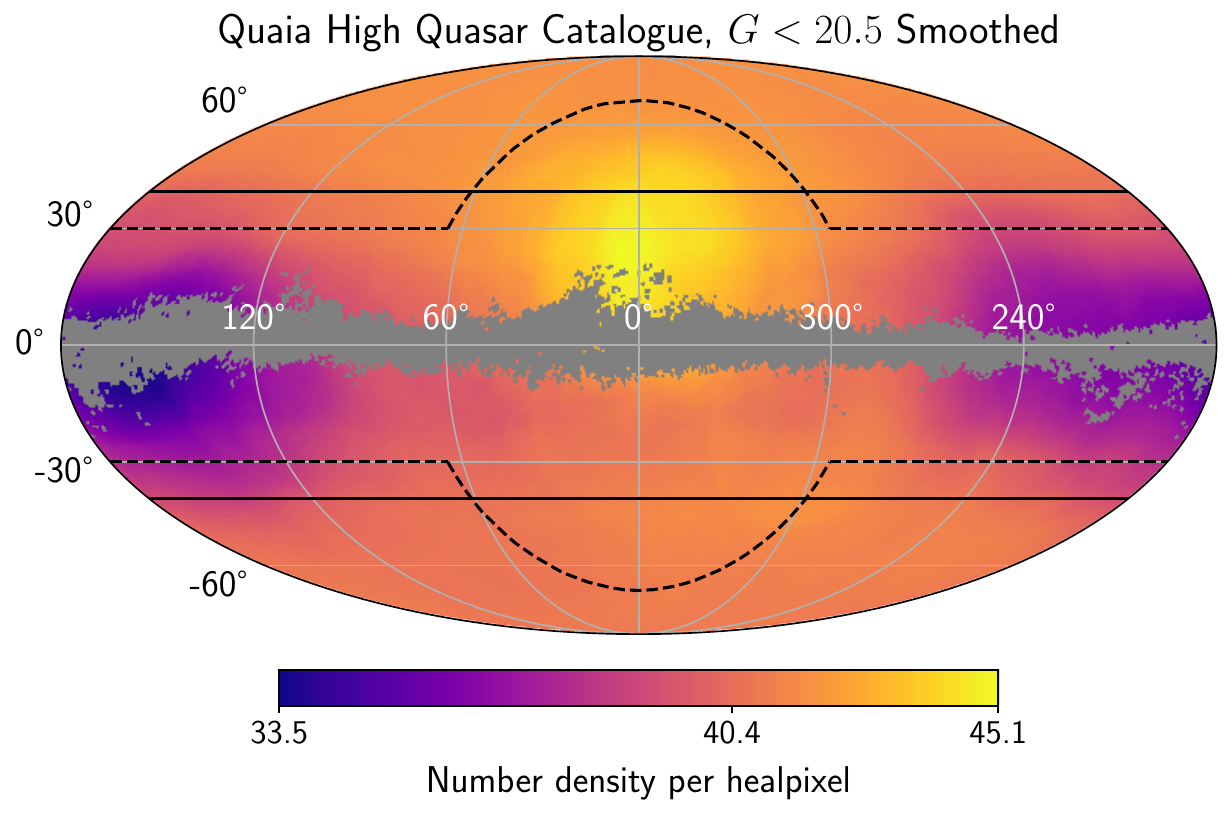}
        \caption{Visualisation of the salient features of the two Quaia catalogues in galactic coordinates, with Quaia low in the left column and Quaia high in the right column. \textit{Top row:} Raw catalogue prior to any additional masking or processing. Note that the catalogue already has an absence of sources near the galactic plane, shown in grey, primarily due to dust absorption. \textit{Middle row:} The selection function provided for both the Quaia catalogues, with the colour scale indicating the probability of source detection associated with each pixel due to factors like dust extinction. \textit{Bottom row:} Both catalogues have been smoothed via a sliding average over a 1 steradian scale after scaling according to the selection function. Deviations in source density along the galactic plane can be seen, with an over-density at the galactic centre and an under-density at mid galactic longitudes. The solid and dashed lines indicate the $40^\circ$ and $30^*$ galactic plane masks respectively.}
        \label{fig:all-maps}
    \end{figure*}

    A visualisation of the raw Quaia low and Quaia high catalogues with number count densities can be seen in the top row of Fig.~\ref{fig:all-maps}.
    These maps, as well as subsequent ones, are displayed in galactic coordinates.
    We show the selection function provided by the Quaia authors for both catalogues in the middle row of Fig.~\ref{fig:all-maps}.
    By visual inspection, dust extinction appears to dominate the map, which explains the dearth of sources near the galactic plane in the raw catalogue.
    Finally, we show smoothed maps in the bottom row of Fig.~\ref{fig:all-maps} for Quaia low and Quaia high.
    To generate the smoothed map, we first scaled the catalogue according to the selection function such that the $i$-th pixel with number of sources $N_i$ (see Section~\ref{subs:binning} for information on how sources are binned) is scaled by $1 / s_i$, where $s_i$ is the value of the selection function at that pixel.
    We then implemented a sliding average; for each pixel, we selected pixels within 1 steradian and computed the mean density.
    These maps give a visual cue of a source over-density near the galactic centre, as well as under-densities near mid galactic longitudes along the galactic plane.
    Superimposed on the maps are two masks we chose to use, which are explained in more detail in Section~\ref{subs:masking}.
    
\section{Approach}
\label{sec:approach}
    \subsection{Catalogue processing}

    \subsubsection{Binning}
    \label{subs:binning}
        In order to prepare the catalogue for analysis, the sky was divided into equal-area pixels using the the pixelisation regime of HEALPix\footnote{\url{https://healpix.sourceforge.io/}} \citep{Gorski2005, Zonca2019} as incorporated in the Python package \textsc{healpy}. 
        $N_\text{side} = 64$ -- generating a total of $49\,152$ pixels -- was chosen, since the selection maps created by the Quaia authors are given at this resolution.
        The choice of $N_\text{side}$ depends upon the fact that for number count analysis, the uncertainty in number counts for each pixel due to shot noise should not be greater than the mean number count for the catalogue.
        We then summed the number of sources within each pixel using their recorded positions in right ascension and declination. 
        This gives a means by which changes in the source density can be discerned as a function of sky position.

    \subsubsection{Masking}
    \label{subs:masking}
        \citet{storeyfisher2023quaia} noted that the selection function is potentially poorly-modelled in the vicinity of the galactic plane.
        In making this judgment, they computed the fractional residuals between a synthetic catalogue generated by randomly sampling over a sphere according to the selection function and the actual Quaia catalogue.
        Around the edge of the plane, the random synthetic catalogue over-predicts the data; additionally, near the galactic centre, the random catalogue seems to under-predict the data.
        We note that in the bottom row of Fig.~\ref{fig:all-maps}, which shows our smoothed map of the Quaia low and Quaia high samples, there indeed appears to be an over-density near $(l,b) \approx (0^\circ, 30^\circ)$ as well as under-densities along the galactic plane from about $l = 120^\circ$ to $l = 240^\circ$. 
        This is in line with the proposition made by the Quaia authors.
        For example, if the galactic centre is under-predicted by the selection function, then $s_i < \tilde{s}_i$ where $\tilde{s}_i$ is some true value of the selection function.
        Thus, in our smoothed maps which originate from scaled number counts, the $i$-th pixel has a number count $N_i / s_i > N_i / \tilde{s}_i$, manifesting as an over-density.
        
        In order to address this issue, we chose to mask the galactic plane with a series of increasingly conservative masks, as the Quaia authors suspected may be necessary at Section 4.5 in \citet{storeyfisher2023quaia}.
        To be explicit, we examined the effect of $|b| < 10^\circ$, $20^\circ$, $30^\circ$ and $40^\circ$ galactic plane masks on the recovered signal in conjunction with an unmasked catalogue. 
        The $30^\circ$ mask curtains much of the problematic regions, but it is still possible that at the edge of the mask the issues at the galactic plane seep into the masked sample.
        Accordingly, in addition to testing with a $40^\circ$ mask, we implemented a circular mask centred on $(l^\circ, b^\circ)=(0,0)$ and subtending a solid angle of $4\,\text{sr}$ in concert with the $30^\circ$ galactic plane mask.
        We denote this as a $30^*$ mask for future reference.
        The $40^\circ$ mask is represented by the solid black line overlaid on the bottom row of Fig.~\ref{fig:all-maps}, and the $30^*$ mask is represented by the dashed black line.
        

    \subsection{Dipole amplitude expectation}
        Since we are ultimately testing the kinematic interpretation of the CMB, we will need to compare the expected dipole amplitude given CMB-inferred motion and the actual recovered dipole from the Quaia sample. 
        Conventionally, this amounts to using equation~\eqref{eq:dipole-magnitude} with $v = v_{\text{CMB}} \approx 369\,\text{km\,s}^{-1}$. 
        This also means that $x$ and $\alpha$ must be ascertained from the sample of galactic sources. 
        Here, we instead use the actual source counts themselves -- rather than their proxy $x$ -- and take the distribution of $\alpha$ to find a distribution of dipole amplitudes $\mathcal{D}$ given $v$.
        This approach is detailed below.

        \subsubsection{Spectral index}
\begin{figure}
                \centering
                \includegraphics[width=\linewidth]{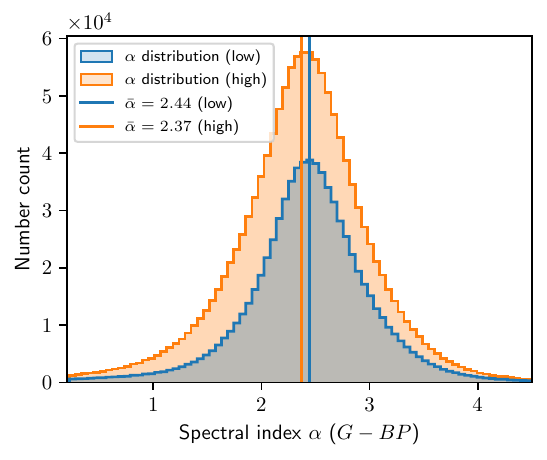}
                \caption{Distribution of spectral indices $\alpha$ in the Quaia low and Quaia high samples computed from $m_{G - BP}$. Blue and orange are used to denote Quaia low and Quaia high respectively. The mean spectral indices $\bar{\alpha}$ for each catalogue are indicated by the vertical lines.}
                \label{fig:alpha-dist-low}
            \end{figure}

            As mentioned earlier, we assume that the $i$-th Quaia source follows a flux power law such that $S_\nu \propto \nu ^ {- \alpha_i}$.
            To find the spectral index $\alpha_i$, we compute the colour magnitude $m_{G - BP}$.
            Since Gaia magnitudes are measured in the Vega system, we use the zero points (ZP) and mean wavelengths of the $G$ and $BP$ bands, as provided in \citep{riello2021}, to determine $\alpha_i$.
            Namely,
            \begin{equation}
                m_\nu = -2.5 \log_{10} S_\nu + \text{ZP}
            \end{equation}
            such that
            \begin{align}
                m_{G - BP} &= 2.5 (\log_{10} S_{BP} - \log_{10} S_{G}) + \text{ZP}_{G} - \text{ZP}_{BP} \\
                \implies \alpha_i &= \frac{k - m_{G - BP}}{2.5 \log_{10} (\nu_{BP} / \nu_{G})} \label{eq:spectral-index}
            \end{align}
            where $k$ is $\text{ZP}_{G} - \text{ZP}_{BP}$ and in the last line we used the assumption that $S_\nu \propto \nu ^ {- \alpha_i}$.
            Equation~\eqref{eq:spectral-index} yields a distribution of spectral indices for Quaia low and Quaia high, which we show in Fig.~\ref{fig:alpha-dist-low}.
            The mean value of $\alpha$ is labelled there only for illustrative purposes; what is important, in our analysis, is the distribution itself.

        \subsubsection{Source number counts}
        \label{subs:source-number-counts}
\begin{figure}
                \centering
                \vspace{5.7mm}
                \includegraphics[width=\linewidth]{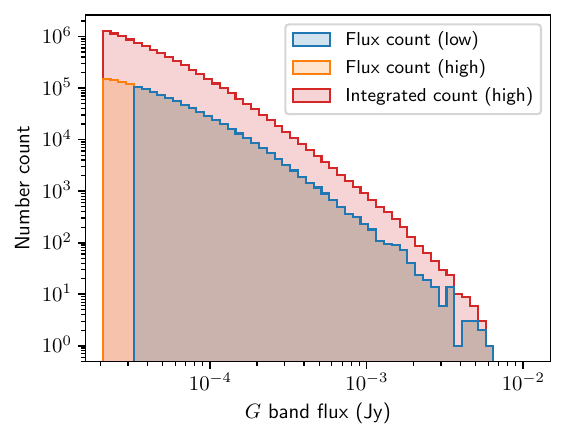}
                \caption{Source number counts binned by $G$ band flux for both Quaia low (blue) and Quaia high (orange). Overlaid in red is the integrated source count above a limiting flux density for Quaia high, which of course includes sources in Quaia low.}
                \label{fig:flux-dist-low}
            \end{figure}

            To find the distribution of fluxes in the Quaia sample, we first convert the $G$ magnitude into a \textit{Gaia} flux using the zero points mentioned above.
            This yields \textit{Gaia} fluxes in units of photoelectrons$\,\text{s}^{-1}$, though we note that these fluxes can also be found by matching each Quaia source with its entry in DR3 by using each entry's \textit{Gaia} DR3 source identifier.
            To convert from these units into Jy, we apply the relevant conversion factor $c_\nu$ found in the \textit{Gaia} documentation \citep{gaiadocumentation}.
            A histogram showing the resultant flux distribution is presented in Fig.~\ref{fig:flux-dist-low}.
            Overlaid there in red is the integrated distribution, i.e. the number of sources above some limiting flux density $N (>S)$. 
            
\begin{figure}
                \centering
                \vspace{4.5mm}
                \includegraphics[width=\linewidth]{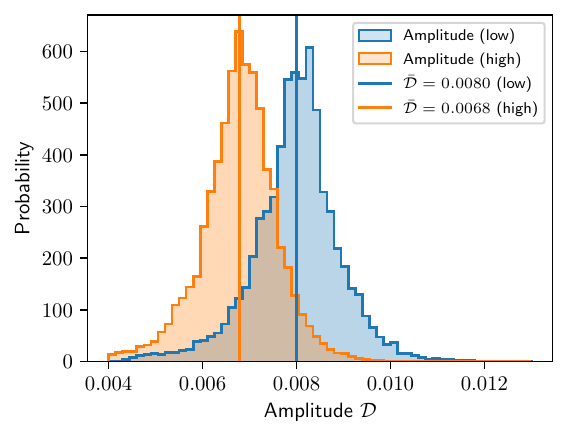}
                \caption{Probability distribution for the dipole amplitude assuming $v_\text{CMB}$. This follows from the analysis in Section~\ref{subs:source-number-counts} and was performed with both the Quaia low and Quaia high samples.}
                \label{fig:CMB-expect-qLow}
            \end{figure}

            In the context of quasar studies, the approach used in describing the source count distribution has been similar to that of radio galaxy studies.
            Specifically, in determining the dipole amplitude, power law fits of the form $S^{-x}$ to the integrated source counts have been used \citep[see e.g.][]{secrest2021}, as well as piece-wise straight line fits, onto which a flux cut is imposed such that the data is constrained to the regime of one of those power laws \citep[see e.g.][]{singal2023}. 
            This traces back to the original conceptual framing of \citet{ellis1984}, which supposed that a radio galaxy population can be described by a power law.
            Yet, by inspection, a straight line fit to the integrated counts in Fig.~\ref{fig:flux-dist-low} is not the best reflection of the actual data.
            In order to get a better hold of the non-linear nature (in logarithmic space) of the data, we instead work directly with the observed fluxes in the $G$ band.
            When we observe a population of sources and their associated fluxes, we expect a number count enhancement in the forward hemisphere and a diminution in the backwards hemisphere; sources become brighter and fainter respectively, and they congregate along the line of motion.
            The conceptual underpinning of \citet{ellis1984} is that the observed source count power law $S^{-x}$ is the resultant of these two effects.
            Thus, in order to find the expected dipole amplitude, the following method was used.
            \begin{enumerate}
                \item We computed the number of sources greater than some limiting flux density $S_0$, denoted as $n_i$.
                \item For the $j$-th source with measured flux $S_j$, a Doppler shift was applied such that $S_j \to S_j \delta^{1 + \alpha}$ where $\delta = \gamma (1 + v \cos \theta)$ and $\gamma$ is the Lorentz factor. This is the relationship between the observed and rest frame flux densities as described in \citet{ellis1984}.
                \item We then computed the number of boosted fluxes greater than some limiting flux density
                and multiplied this sum by $\delta^2$, which anticipates relativistic aberration. We denote this final value as $n_b$.
                \item Combining the above, we then calculated the expected dipole amplitude as
                \begin{equation}
                    \mathcal{D} = \frac{n_b - n_i}{n_i}.
                \end{equation} 
            \end{enumerate}
            Note that we have selected units where $c = 1$, and if we take a measurement along the line of motion -- the direction of maximal density enhancement -- then $\theta = 0$ so $\delta = \gamma (1 + v)$.
            We also fixed $S_0$ to be near the flux limit of the catalogue. 
            For Quaia low, $G = 20.0$ corresponds to a flux of $\approx 3.27 \times 10^{-5}\,\text{Jy}$ in the $G$ band, so we took $S_0 = 3.3 \times 10^{-5}\,\text{Jy}$.
            For Quaia high, $G = 20.5$ corresponds to $\approx 2.06 \times 10^{-5}\,\text{Jy}$, and so we used $S_0 = 2.1 \times 10^{-5}\,\text{Jy}$.
            Then, substituting $v = v_\text{CMB}$ into the above analysis and randomly sampling $\alpha$ from Fig.~\ref{fig:alpha-dist-low} $50\,000$ times, we find a dipole described by the distributions in Fig.~\ref{fig:CMB-expect-qLow} with mean amplitude $\mathcal{\bar{D}} \approx 0.0080$ for Quaia low and $\bar{\mathcal{D}} \approx 0.0068$ for Quaia high.
        
        \subsection{Bayesian analysis}
        \label{sub:bayesian-methods}    
        \subsubsection{Bayes's theorem}
            With the Quaia sample prepared for analysis, we now explore how the framework of Bayesian statistics provides a natural language to test competing hypotheses and understand the Quaia data.
            The specific models we consider are explained in Section~\ref{sub:hypotheses-considered}, but for now, a Bayesian approach to model comparison can be broken down into two key steps or levels \citep{mackay2003}.
            At the first level of inference, a model's parameters are optimised and the posterior distributions for those parameters are recovered.
            This amounts to solving Bayes's theorem, where
            \begin{equation}
                P( \mathbf{\Theta} | \mathbf{D}, M ) = \frac{\mathcal{L}(\mathbf{D} | \mathbf{\Theta}, M) \pi(\mathbf{\Theta} | M)}{\mathcal{Z}(\mathbf{D} | M)}. \label{eq:bayes-theorem}
            \end{equation}
            We have recast the notation of Bayes's theorem in line with \citet{speagle2020} to better indicate what each term means in the context of model inference; namely, $\mathbf{D}$ refers to the data and $\mathbf{\Theta}$ refers to the set of parameters pertaining to model $M$.
            $\mathcal{L}$, $\pi$ and $\mathcal{Z}$ refer to the likelihood, prior and evidence or marginal likelihood functions respectively, and $P$ denotes the resulting posterior probability distribution.
        

        \subsubsection{Evidence and the Bayes factor}
        \label{subs:evidence}
            At the first level, the marginal likelihood only represents a normalisation term.
            However, at the second level of inference, competing hypotheses or models are ranked using the Bayes factor, which is the ratio of the marginal likelihoods for each model.
            Specifically, if we wish to determine the relative levels of support for models $M_1$ and $M_2$, we may compute
            \begin{equation}
                \ln B_{12} = \ln \mathcal{Z}_1 - \ln \mathcal{Z}_2
            \end{equation}    
            where $B$ denotes the Bayes factor.
            It is advantageous to work with natural logarithms since the actual value of the marginal likelihood is generally very small.
            In this work, any quoted marginal likelihood or Bayes factor is a natural logarithm.
            The interpretation of the level of support depends on the value of the Bayes factor.
            In this example, $\ln B_{12} > 0$ means that $M_1$ is preferred over $M_2$, with larger values indicating more support for $M_1$.
            For reference, \citet{kass1995} provide a qualitative description depending on the value of the Bayes factor, although we stress that this is not meant to be definitive.
            We also note that the table in their work has units of $2 \ln B$.
            We present our values as $\ln B$, so the reader will need to keep the factor of two in mind when comparing our values to the ranges appearing in \citet{kass1995}.

            Since the marginal likelihood is an integral over all parameter space $\mathcal{Z} = \int_{\Omega_{\mathbf{\Theta}}} \mathcal{L}(\mathbf{\Theta}) \times \pi(\mathbf{\Theta}) \, d \mathbf{\Theta}$, models with excessive parameters which waste parameter space are intrinsically disfavoured.
            To a rough approximation, the marginal likelihood can be written as
            \begin{equation}
                \mathcal{Z} \approx \mathcal{L}(\mathbf{\Theta}_{\text{MP}}) \frac{\Delta \mathbf{\Theta}}{\Delta^0 \mathbf{\Theta}}
            \end{equation}
            following the argument in \citet{mackay2003}. $\mathcal{L}(\mathbf{\Theta}_{\text{MP}})$ is the value of the likelihood function at the most optimal set of parameters, where $\Delta \mathbf{\Theta} / \Delta^0 \mathbf{\Theta}$ can be analogised as the ratio of the peak in the likelihood function to the width of the prior distribution.
            This term is called the Occam factor, and it generally penalises models which squander parameter space and only have explanatory power (i.e. a high likelihood) for a comparatively small region of parameter space.
            
            For each of our hypotheses or models, we compute the marginal likelihoods and Bayes factors, using these metrics to rank them.
            The result of this process is a quantitative evaluation of which hypothesis has the strongest level of support, and as such what kind of model best accounts for the Quaia sample.
    
        \subsubsection{Nested sampling}
            \label{subs:nested-sampling}
            A key difficulty with Bayesian methods is that the marginal likelihood cannot usually be computed analytically, and is somewhat expensive to determine numerically.
            However, modern computational methods are well-adapted to this challenge, and provide efficient algorithms which are easy to implement.
            In this work, we take advantage of the Nested Sampling (NS) algorithm \citep{skilling2004, skilling2006}.
            The focus of NS is on first determining the marginal likelihood and then evaluating the posterior distributions for a model's parameters as a `subsidiary element' \citep{skilling2004}.
            In trying to evaluate the marginal likelihood, which is an integral over all parameter space as mentioned earlier, NS recasts the integral to be over the prior instead and generates iso-likelihood contours or shells of increasing likelihood \citep{speagle2020}.
            This gives an effective means of evaluating $\mathcal{Z}$ and an associated uncertainty, with the posterior distribution as a byproduct.
            Here, we have used \textsc{Dynesty}\footnote{\url{https://pypi.org/project/dynesty/}} \citep{dynesty-v2.1.2}, a Python package which implements the NS algorithm.
        

        \subsection{Likelihood functions}
        \label{sub:likelihood-functions}
            Armed with the methods of Bayesian analysis, the next step in our approach is to construct likelihood functions $\mathcal{L}$ to be placed in Bayes's theorem at equation~\eqref{eq:bayes-theorem}.
            There are essentially two approaches that we can apply; these each test the same underlying assumption, but are framed in slightly different ways.
            We describe each of these in a model-invariant manner, eluding to how they would be altered to fit a particular hypothesis along the way.

            \subsubsection{Poissonian statistics}
                In the first approach, the binning process referenced in Section~\ref{subs:binning} is analogous to a Poisson point process, and variations in number density (occupancy) across each pixel can ideally be explained by two factors: shot noise and an intrinsic signal.
                Shot noise necessitates that we associate each pixel with a Poisson distribution, i.e. the occupancy for a given pixel is a random variate drawn from a Poisson distribution.
                The intrinsic signal could be a dipole, where sources in the forward hemisphere are associated with higher number densities than sources in the backward hemisphere.
                This modifies the rate parameter of the Poisson distribution describing a pixel's number density.

                In light of this, the probability $P$ of observing $N_i$ sources in pixel $i$ can be written as
                \begin{equation}
                    P(N_i | \, \mathbf{\hat{p}_i} ) = \frac{\lambda_i^{N_i} e^{-\lambda_i}}{N_i!}
                \end{equation}
                where $\mathbf{\hat{p}_i}$ is a unit vector pointing towards the $i$-th pixel and $\lambda_i$ is the rate parameter for the $i$-th pixel.
                The expected number of occupants in a pixel is just the rate parameter: $E [N_i (\mathbf{\hat{p}_i})] = \lambda_i$.

                In practice, as explained in Section~\ref{sec:quaia-catalogue}, Quaia is packaged with a selection function that ascribes each pixel with a probability.
                On top of statistical fluctuations and variations due to the underlying signal, pixels might have source under-densities because of factors like extinction from dust in the galactic plane.
                This needs to be accounted for insofar that we are examining the assumption of homogeneity and isotropy.
                We can do this through attenuating the rate parameter by the value of the selection function $s_i$ (a probability between 0 and 1) at the $i$-th pixel.
                Explicitly, this means that
                \begin{equation}
                    P(N_i | \, \mathbf{\hat{p}_i} ) = \frac{(\lambda_i \times s_i)^{N_i} e^{-\lambda_i \times s_i}}{N_i!} \label{eq:poisson-prob}
                \end{equation}
                such that $E [N_i (\mathbf{\hat{p}_i})] = \lambda_i \times s_i$.

                By associating each pixel on the sky with a probability determined from Poisson distributions, the likelihood function can be written as the product of all probabilities.
                As a logarithm, this becomes
                \begin{equation}
                    \ln \mathcal{L} = \sum_{i=1}^{n_{\text{pix.}}} \ln P(N_i | \, \mathbf{\Theta} ) \label{eq:poisson-likelihood}
                \end{equation}
                for total number of pixels $n_{\text{pix.}}$.
                
            \subsubsection{Point-by-point analysis}
                Turning to the second approach, we are not restricted to our choice of Poissonian statistics.
                We can see this by examining each source individually such that the likelihood function is now the product of all points, not the product of all pixels. 
                Thus, we term this the `point-by-point' approach.
                However, the sky is still discretised to simplify the calculation; points within a certain pixel are assumed to have the same probability.
                This approach was adopted by \citet{conn2011,conn2012} in their distance determinations using the tip of the red giant branch for sparsely populated systems.

                Let us introduce the function $f_i$, which describes the anticipated signal at pixel $i$.
                This is a model-dependent term, the functional form of which we leave for Section~\ref{sub:hypotheses-considered}.
                If we examine each source individually and not as a member of a pixel, then the distribution associating points with a probability depending on their position on the sky takes the form $f_i$, at least up to a normalising constant.
                Thus, the contribution to the likelihood function at the $i$-th pixel is $\propto (s_i \times f_i)^{N_i}$, since we are taking the product over all points $N_i$ in the pixel.

                To normalise this distribution for the $i$-th pixel (denoted as $\hat{f}_i$ below), which is critical after application of a mask and selection map, we sum over all unmasked pixels such that
                \begin{equation}
                    \hat{f}_i = \frac{s_i \times f_i}{\sum_{i=1}^{n_{\text{pix.}}} s_i \times f_i}. \label{eq:hybrid-prob}
                \end{equation}
                Thus, the natural logarithm of the likelihood function can be written as
                \begin{equation}
                    \ln \mathcal{L} = \sum_{i=1}^{n_{\text{pix.}}} N_i \ln \hat{f}_i. \label{eq:hybrid-likelihood}
                \end{equation}
                In principle, both the Poissonian and point-by-point models should give consistent results, since both represent slightly different approaches to describe the same underlying effect.
                We confirm this in our results at Section~\ref{sec:results}.
    
    \subsection{Hypotheses under consideration}
    \label{sub:hypotheses-considered}
    \subsubsection{$M_0$: Monopole (Null)}
        Suppose that the distribution of sources is in fact homogeneous and isotropic in the observer's frame. 
        Then a monopole signal is anticipated, where pixels are expected to have some mean number density $\bar{N}$ irrespective of the location of the pixel on the sky.
        Expressed differently, the expected number density is
        \begin{equation}
            E [ N_i(\mathbf{\hat{p}_i}) ] = \bar{N}.
        \end{equation}
        This acts as the null hypothesis for our study.
        In order to compute the likelihood function in the Poissonian case, the rate parameter $\lambda_i$ of equation~\eqref{eq:poisson-prob} is replaced with $\bar{N}$, whereas in the point-by-point case $f_i$ is set to 1.
        These are then substituted into equations~\eqref{eq:poisson-likelihood} and \eqref{eq:hybrid-likelihood} respectively. 
        Thus, $\mathbf{\Theta}_{\text{Pois.}} = \{ \bar{N} \}$ in the Poissonian case and $\mathbf{\Theta}_{\text{P}\times\text{P}} = \emptyset$ in the point-by-point case.

    \subsubsection{$M_1$: Dipole}
        As an alternative to the null hypothesis, we introduce the vector $\mathbf{D}$, which points in the direction of the dipole signal and has amplitude equal to the magnitude of the dipole (see e.g. equation~\eqref{eq:dipole-magnitude}). 
        The anticipated number count is then described by the sum of the monopole and dipole signals \citep[see e.g.][]{dam2023} such that
        \begin{equation}
            E [N_i (\mathbf{\hat{p}_i}) ] = \bar{N} + \bar{N} (\mathbf{D} \cdot \mathbf{\hat{p}_i}) = \bar{N}(1 + \mathcal{D} \cos \theta_i) \label{eq:dipole-expectation}
        \end{equation}
        where $\theta_i$ is the angle between the dipole direction and the $i$-th pixel vector, and $\mathcal{D}$ is the magnitude of the dipole.
        
        Here, Equation~\eqref{eq:dipole-expectation} is the rate parameter $\lambda_i$ that is inserted into equation~\eqref{eq:poisson-prob} where Poissonian statistics is used, and $f_i = 1 + \mathcal{D} \cos \theta_i$ in the point-by-point analysis for the purposes of equation~\eqref{eq:hybrid-prob}.
        Evidently, the parameter spaces are given by $\mathbf{\Theta}_{\text{Pois.}} = \{ \bar{N}, \mathcal{D}, l, b \}$ and $\mathbf{\Theta}_{\text{P}\times\text{P}} = \{ \mathcal{D}, l, b \}$, where $l$ and $b$ characterise the direction of the dipole in galactic coordinates.

    \subsubsection{$M_2$: Double dipole}
        \label{subs:double-dipole}
        The presence of an over-density region just above the galactic center and other under-densities along the galactic plane in both Quaia low and Quaia high, which we described in Section~\ref{subs:masking}, hints towards the fact that the net dipole in Quaia might be a combination of two dipoles.
        The net dipole modulation is then the multiplication of two individual dipoles, and the expected number density for the $i$-th pixel is given by
        \begin{equation}
            E [N_i (\mathbf{\hat{p}_i}) ] = \bar{N}\left[(1 + \mathcal{\bf D_{1}} \cdot {\bf \hat{p}_{i}}) \times (1 + \mathcal{\bf D_{2}} \cdot {\bf \hat{p}_{i}})\right] \label{eq:double-dipole-multiplicative}
        \end{equation}
        This is used as a rate parameter for equation~\eqref{eq:poisson-prob} where Poissonian statistics is used, and $f_i = (1 + \mathcal{\bf D_{1}} \cdot {\bf \hat{p}_{i}}) \times (1 + \mathcal{\bf D_{2}} \cdot {\bf \hat{p}_{i}})$ in the point-by-point case for equation~\eqref{eq:hybrid-prob}.
        Thus, our parameter space is $\mathbf{\Theta}_{\text{Pois.}} = \{ \bar{N}, \mathcal{D}_{1}, l_{1}, b_{1}, \mathcal{D}_{2}, l_{2}, b_{2} \}$ and $\mathbf{\Theta}_{\text{P}\times\text{P}} = \{ \mathcal{D}_{1}, l_{1}, b_{1}, \mathcal{D}_{2}, l_{2}, b_{2} \}$.
        
        The assumption here is that the two dipoles were generated at different times and due to different factors. 
        So, the observer's frame was already anisotropic due to one dipole at the genesis of the second dipole, and it makes sense to apply an extra modulation on top of an already modulated sky.
        If this was not the case and the genesis time were the same for the two, then the net motion should be in a direction that is in between the two dipoles and hence only a single dipole would be observed. 
        Even if such a scenario seems unlikely, it is worth examining as the marginal likelihood will balance the explanatory power of this model and its complexity, potentially revealing deeper insight into the nature of the Quaia sample.
        
    
    
    \subsubsection{$M_3$: Quadrupole}
        For the sake of completeness, we also test for an underlying quadrupole signal. We postulate that this signal is $\propto \cos^2 \theta$, such that the expected number density for the $i$-th pixel is
        \begin{equation}
            E[N_i (\mathbf{\hat{p}_i}) ] = \bar{N} (1 + \mathcal{\tilde{D}} \cos^2 \theta_i).\label{eq:quadrupole-expectation}
        \end{equation}
        Our parameter spaces are $\mathbf{\Theta}_{\text{Pois}.} = \{ \bar{N}, \mathcal{\tilde{D}}, l, b \}$, where the tilde on the signal magnitude suggests the fact that the quadrupole amplitude is not necessarily the same as the foregoing dipole models. For the point-by-point case, $\mathbf{\Theta}_{\text{P}\times\text{P}} = \{ \mathcal{\tilde{D}}, l, b \}$. 
        The expected form of this signal is inserted into equations~\eqref{eq:poisson-prob} and \eqref{eq:hybrid-prob}, as already described above. 
        It is worth noting that the quadrupole model is in fact a special case for the double dipole model with $\mathcal{\bf D_{1}}\cdot {\bf \hat{p}_{i}} = -\mathcal{\bf D_{2}}\cdot {\bf \hat{p}_{i}}$ in equation~\eqref{eq:double-dipole-multiplicative}. 
        
        
    \subsubsection{$M_4$: Dipole pointing towards the kinematic dipole}
        \label{subs:amplitude}
        In order to verify the kinematic interpretation of the cosmic dipole, it is worth fixing certain parameters to their value as determined from the CMB dipole while varying the others.
        Doing so is advantageous, as the marginal likelihood for a kinematic hypothesis can be compared to models where such an interpretation is not assumed.
        
        Here, we fix the model's dipole to the direction of the CMB dipole, namely $(l,b) = (264\dotdeg0.21, 48\dotdeg253)$, and leave the amplitude as a free parameter. 
        The expectation of the number density is identical to equation~\eqref{eq:dipole-expectation} except with $\theta$ fixed to $\theta_\text{CMB}$, and so the parameters pertaining to each model are $\mathbf{\Theta}_{\text{Pois.}} = \{ \bar{N}, \mathcal{D} \}$ and $\mathbf{\Theta}_{\text{P}\times\text{P}} = \{ \mathcal{D} \}$.
        
    \subsubsection{$M_5$: Dipole fixed by the kinematic dipole velocity}
        \label{subs:direction}
        Another test involves fixing the magnitude of the dipole to the CMB value while allowing its direction to vary. 
        The dipole magnitude is fixed to the mean values of $\mathcal{D}$ calculated using the method described in \ref{subs:source-number-counts}.
        More explicitly,  $\mathcal{D} =  0.0080$ for Quaia low and $\mathcal{D} =  0.0068$ for Quaia high.
        Here, equation~\eqref{eq:dipole-expectation} applies with $\mathcal{D}$ fixed, and so the parameters for each model are $\mathbf{\Theta}_{\text{Pois.}} = \{ \bar{N}, l, b \}$ and $\mathbf{\Theta}_{\text{P}\times\text{P}} = \{ l, b \}$.

    \subsubsection{$M_6$: CMB motion}
        \label{subs:cmb-motion}
        Finally, we may totally align this model's dipole in both direction and magnitude with the CMB kinematic dipole and compute the marginal likelihood. 
        This gives a metric by which the veracity of CMB-aligned motion in the Quaia sample compares to an inferred direction after parameter optimisation.
        The parameters in this case are $\mathbf{\Theta}_{\text{Pois.}} = \{ \bar{N} \}$ and $\mathbf{\Theta}_{\text{P}\times\text{P}} = \emptyset$.

    \subsection{Choice of priors}
        As a final step, we determine the prior functions $\pi (\mathbf{\Theta} | M)$ for equation \eqref{eq:bayes-theorem}, which is needed for each model's parameters.
        The choice of prior represents our belief about what values the parameters are likely to take before knowledge of the data.
        \begin{itemize}
            \item We adopted a broad prior for the dipole amplitude $D$, as well as the double dipole amplitudes $D_1, D_2$, choosing them from a uniform distribution $D, D_1, D_2 \sim \mathcal{U}[0, 1]$.
            This choice was motivated by the significant uncertainty in the magnitude of the dipole across a diverse spectrum of independent tests \citep[see e.g.][]{snowmass2022, aluri2023}.
            In contrast, we sampled the quadrupole amplitude according to $\tilde{D} \sim \mathcal{U}[-1,0]$, since in testing we found that the positive amplitude solution restricted the solution to a poorer fit at the north galactic pole.
            \item For the direction parameters, in internal calculations we work in equatorial coordinates but convert the posterior afterwards to galactic coordinates for presentation.
            We denote these equatorial coordinates ($\phi$, $\theta$) for right ascension $\phi$ in radians and co-declination $\theta$ in radians.
            For the dipole direction, we uniformly sampled over the surface of a unit sphere, and so $\phi \sim \mathcal{U}[0, 2\pi]$ and $\theta \sim \cos^{-1} (1 - 2 u)$ where $u \sim \mathcal{U}[0,1]$.
            For the double dipole directions, we took $\phi_1, \phi_2 \sim [3 \pi /2, 5 \pi /2], [\pi / 2, 3 \pi / 2]$ and $\theta_1, \theta_2 \sim \cos^{-1} (1 - 2u)$ to prevent cross-talk between the two signals in the posterior distribution at opposite hemispheres.
            In a similar sense, for the quadrupole model we took $\phi \sim [\pi / 2, 3 \pi / 2]$ and $\theta \sim \cos^{-1} (1 - 2u)$, since there is a degenerate solution in the other hemisphere.
            \item For the mean number density or monopole signal $\bar{N}$, we took $\bar{N} \sim \mathcal{U}[0,30]$ for Quaia low and $\bar{N} \sim \mathcal{U}[0,50]$ for Quaia high.
        \end{itemize}
    
\section{Results}
\label{sec:results}
    With our models outlined, we now turn to the recovered parameters and marginal likelihoods of each model.
    Since we are chiefly interested in this comparative assessment, and not the actual value of the evidence itself, an easy way to represent the data is by computing the Bayes factor for model $M_i$ with respect to the null hypothesis $M_0$: this is $\ln B_{i0} = \ln \mathcal{Z}_i - \ln \mathcal{Z}_0$.
    All models are being assessed with a common benchmark -- the null hypothesis -- which allows a quick identification of the strongest hypothesis.
    Thus, the relative level of support for one hypothesis ($M_i$) over another hypothesis ($M_j$) is simply computed by
        \begin{align}
            \ln B_{ij} = \ln \mathcal{Z}_i - \ln \mathcal{Z}_j &= (\ln \mathcal{Z}_i - \ln \mathcal{Z}_0 ) - (\ln \mathcal{Z}_j - \ln \mathcal{Z}_0) \\
            &= \ln B_{i0} - \ln B_{j0}.
        \end{align}
    One point to keep in mind, however, is that since our Poissonian and point-by-point approaches of Section~\ref{sub:likelihood-functions} use different likelihood functions, the actual value of $\mathcal{Z}$ for a given model is very different across the two approaches.
    What should not change appreciably between them, as we show below, is the Bayes factor.
    
    Since our experiment has many variations -- namely different models, masks, approaches and catalogues -- we generated a considerable number of marginal likelihoods and hence Bayes factors, with the latter being tabulated in Appendix~\ref{sec:appendix}.
    There, the highlighted cells draw attention to the model with the highest Bayes factor for a given galactic mask.
    These are too voluminous to give in their entirety here.
    Certain salient results, however, are referenced periodically in the following text to substantiate our findings.
    \subsection{Quaia low}

    \subsubsection{Low galactic masks: $|b| < 10^\circ$, $20^\circ$, $30^\circ$}
        \label{subs:low-gal-masks-low}
        In this masking regime, the prevailing model is the double dipole ($M_2$).
        This is exemplified by the Bayes factors in in Table~\ref{tab:bayes-low-p2p} and Table~\ref{tab:bayes-low-poisson} for the point-by-point and Poissonian approaches respectively. 
        Note here that the conclusions -- that is, the level of support for each model -- are the same across both approaches.
        
        As an example, we placed the Bayes factors for a $30^\circ$ mask with Quaia low and the point-by-point method in Table~\ref{tab:data-qlow-c30-p2p}.
        \begin{table}
        \centering
            \begin{tabular}{l S[table-format=7.1, round-precision=1] S[table-format=2.1,round-precision=1]}
            \hline
             Model (Point-by-point)                             & {$\ln B_{i0}$} \\ \hline 
             $M_0$ (Null)                               & \hspace{10mm}\textemdash    \\
             $M_1$ (Dipole)                             & 14.4           \\
             \rowcolor{black!10} $M_2$ (Double Dipole)  & 18.9           \\ 
             $M_3$ (Quadrupole)                         & 6.2            \\
             $M_4$ (Kinematic Direction)                & 12.776030      \\
             $M_5$ (Kinematic Velocity)                 & 13.99662       \\
             $M_6$ (Kinematic Dipole)                   & 15.494468      \\\hline
            \end{tabular}
        \caption{Table of Bayes factors by model for a $30^\circ$ mask with the Quaia low sample and using the point-by-point approach. The highlighted cell represents the model with the highest Bayes factor, indicating it has the strongest level of support.}
        \label{tab:data-qlow-c30-p2p}
        \end{table}
        The level of support of the double dipole ($M_2$) over the dipole ($M_1$) is $\ln B_{21} = 4.5$, which represents strong support.
        The model with the second-highest Bayes factor in Table~\ref{tab:data-qlow-c30-p2p} -- the kinematic dipole $M_6$ -- yields a relative Bayes factor of $\ln B_{26} = 3.4$.
        For the $|b| < 30^\circ$ mask, this is the closest a competing model comes to out-competing the double dipole.
        As we move towards less conservative galactic plane masks, this difference in general increases.
        For $|b| < 10^\circ, 20^\circ$, the dipole model ($M_1$) has the second-highest Bayes factor as seen in Tables \ref{tab:bayes-low-p2p} and \ref{tab:bayes-low-poisson}, followed by the kinematic velocity model ($M_5$).
    
    \subsubsection{High galactic mask: $|b| < 40^\circ$ and $30^*$}
        \label{subs:high-gal-masks-low}
        For the $|b| < 40^\circ$ galactic plane mask, no longer is the double dipole the prevailing model.
        Instead, the kinematic dipole $M_6$ has the highest Bayes factor, followed by $M_5$ (kinematic velocity) and $M_4$ (kinematic direction).
        Recall that model $M_6$ assumes a dipole totally aligned with the CMB dipole and possessing the same amplitude.
        The dominance of model $M_6$ is consistent across both Poissonian and point-by-point approaches, as expected.
        
        Critically, this sheds light on the fact that a transition is occurring from the $|b| < 30^\circ$ mask to the $|b| < 40^\circ$ mask; the support for the fitted hypotheses $M_1$--$M_3$ dwindles substantially, while the kinematic hypotheses $M_4$--$M_6$, which generally have less parameters, gain comparatively higher Bayes factors.
        This is also evinced by the $30^*$ mask, which incorporates both the $|b| < 30^\circ$ and a $4 \text{ sr}$ circular mask centred on $(l,b) = (0^\circ, 0^\circ)$.
        The double dipole $M_2$ and kinematic dipole $M_6$ have comparable Bayes factors in this regime, and so the $30^*$ mask must represent an intermediate stage of this transition.
    
    \subsection{Quaia high}
        
        \subsubsection{Low galactic masks: $|b| < 10^\circ, 20^\circ, 30^\circ$}
        \label{subs:low-gal-masks-high}
            Similar to Quaia low, the prevailing model is the double dipole ($M_2$).
            This is again exemplified by the Bayes factors in Tables~\ref{tab:bayes-high-p2p} and \ref{tab:bayes-high-poisson} for the point-by-point and Poissonian approaches respectively.
            The level of support is the same across these approaches.

            The Bayes factors for the $30^\circ$ mask with the point-by-point method are reproduced in Table~\ref{tab:data-qhigh-c30-p2p} for reference.
            \begin{table}
            \centering
                \begin{tabular}{l S[table-format=7.1, round-precision=1] S[table-format=2.1,round-precision=1]}
                    \hline
                 Model (Point-by-point)                            & {$\ln B_{i0}$}\\ \hline
                 $M_0$ (Null)                              & \hspace{10mm}\textemdash   \\
                 \rowcolor{black!10} $M_1$ (Dipole)        & 49.6          \\
                 \rowcolor{black!10} $M_2$ (Double Dipole) & 49.6          \\ 
                 $M_3$ (Quadrupole)                        & 1.6           \\
                 $M_4$ (Kinematic Direction)               & 29.2          \\
                 $M_5$ (Kinematic Velocity)                & 35.3          \\
                 $M_6$ (Kinematic Dipole)                  & 28.7          \\\hline
                 \end{tabular}
            \caption{Table of Bayes factors by model for a $30^\circ$ mask with the Quaia high sample and using the point-by-point approach. The highlighted cell represents the model with the highest Bayes factor, indicating it has the strongest level of support.}
            \label{tab:data-qhigh-c30-p2p}
            \end{table}
            Curiously, there is equal support for both the double dipole ($M_2$) and the dipole ($M_1$) with $\ln B_{21} = 0$, although for the Poissonian approach the Bayes factor for $M_2$ is slightly higher than $M_1$ (see Tables~\ref{tab:bayes-high-p2p} and \ref{tab:bayes-high-poisson}).
            This difference, however, does not change the interpretation significantly since it at best suggests a marginal level of support for $M_2$ over $M_1$. 
            Further, the model with the third-highest Bayes factor in Table~\ref{tab:data-qhigh-c30-p2p} -- the kinematic velocity model $M_5$ -- yields a relative Bayes factor of $\ln B_{25} = 14.3$.
            This suggests overwhelming support for both the double dipole and dipole models at $|b| < 30^\circ $ mask over the other competing explanations.
            
            For lower galactic mask angles, the double dipole model is consistently favoured over the dipole model, with the difference in Bayes factors increasing as the galactic mask angle $b$ decreases.

        \subsubsection{High galactic mask: $|b| < 40^\circ$}
            \label{subs:high-gal-masks-high}
            Unlike the Quaia low sample, $M_4$ (kinematic direction) is the prevailing explanation with a $|b| < 40^\circ$ galactic plane mask on Quaia high.
            This is consistent across both the point-by-point and Poissonian approaches, as expected.
            This being said, the Bayes factors (see Tables~\ref{tab:bayes-high-p2p} and \ref{tab:bayes-high-poisson}) for all models  except $M_3$ and $M_0$ are comparable, with the difference between the lowest and highest Bayes factor being $\approx 1$ log unit.
            This suggests that each model is on a similar footing in terms of explanatory power; one model does not totally dominate over the others.
            At best, the largest Bayes factor $\ln B_{41} \approx 1$ suggests some positive support for $M_4$ over the double dipole ($M_1$).

            Interestingly, with the $30^*$ mask, the dipole ($M_1$) is the favoured hypothesis, and the kinematic dipole ($M_6$) offers the next-best explanation of the data.
            The difference between the Bayes factor for $M_1$ and the other models is in general higher than the differences between $M_4$ and the other hypotheses with the $40^\circ$ mask, suggesting the dipole $M_1$ is more dominant in this regime.
            

            

\section{Discussion \& Conclusions}\label{sec:discussion}
    \subsection{Prevailing double dipole as the effect of over-densities}
    \label{sub:prevailing_double_dipole}
    We turn first to explaining the dominance of the double dipole at low galactic masks as opposed to the other hypotheses, which is observed in both Quaia low and Quaia high.
    This was outlined in Sections~\ref{subs:low-gal-masks-low} and \ref{subs:low-gal-masks-high}.
    In essence, what must be explained is why the double dipole -- despite having a more significant penalty from the Occam factor for its complexity -- can provide a comparatively better explanation of the data than the other models.
    
\begin{figure}
            \centering
            \includegraphics[width=\linewidth]{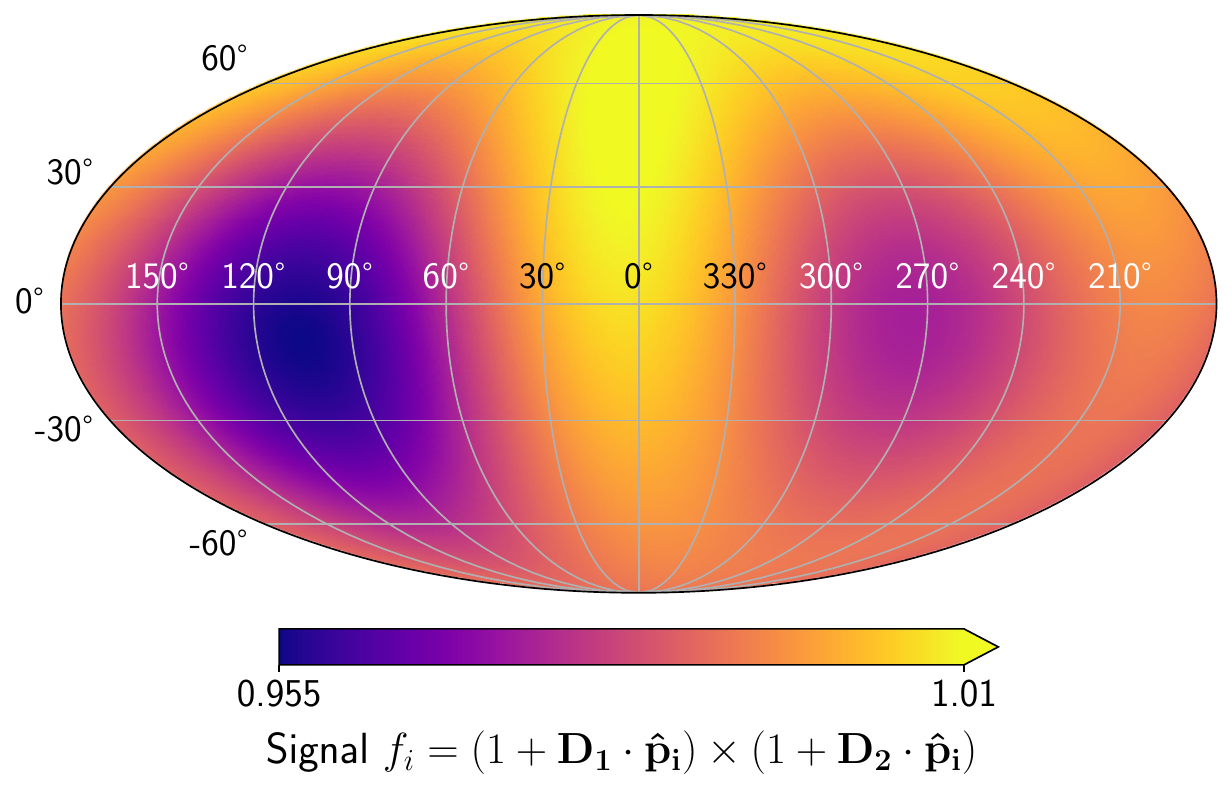}
            \hfill
            \includegraphics[width=\linewidth]{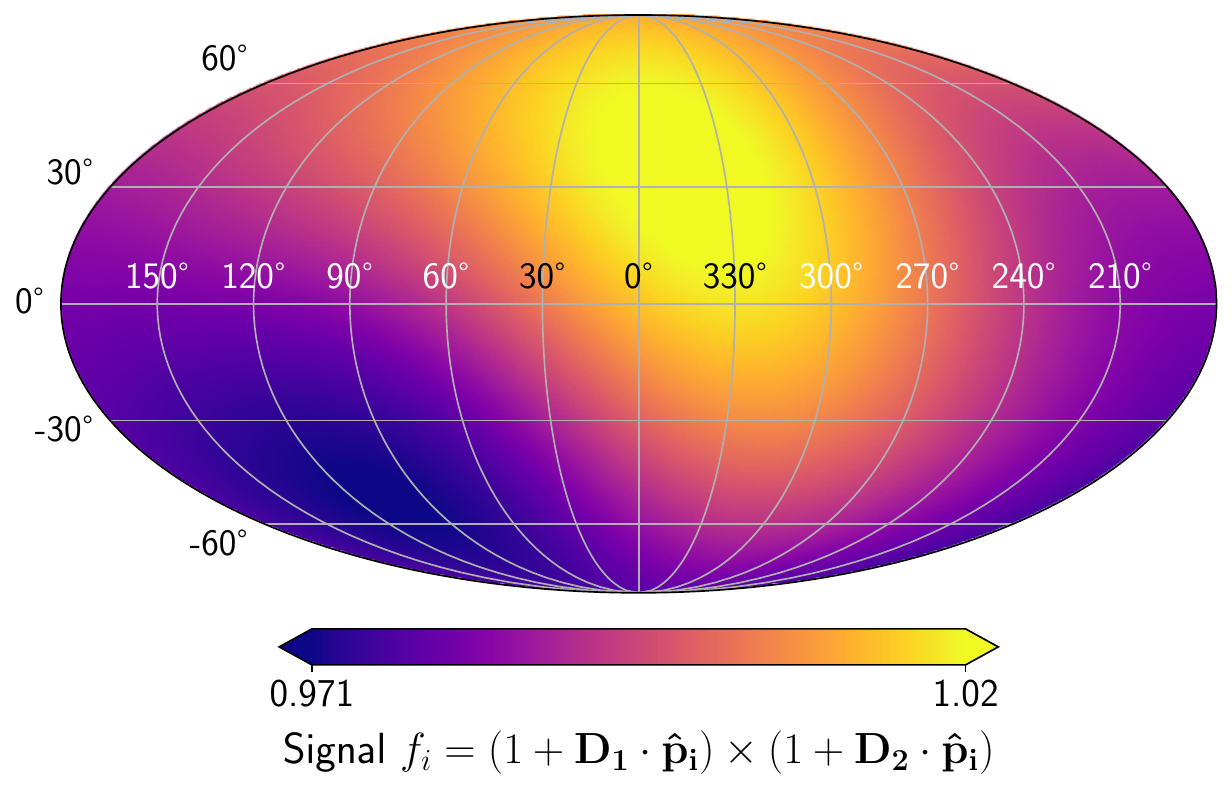}
            \caption{Reconstructed signal using the best fit parameters for model $M_2$ (double dipole) tested on a $30^\circ$ mask. The functional form of this signal is as explained in Section~\ref{subs:double-dipole}. By visual inspection, the nature of this signal is similar to the pattern of over-densities and under-densities shown in the bottom row of Fig.~\ref{fig:all-maps}, since it is this pattern that the model is trying to fit. \textit{Top:} Fit with Quaia low ($G<20.0$). \textit{Bottom:} Fit with Quaia high ($G<20.5$).}
            \label{fig:signal-MultDipHyb-low}
        \end{figure}

    To better see how the model is fitting the data, we extracted the single best fit values (highest marginal probability) for the double dipole with a $30^\circ$ mask and computed the signal term $f_i = (1 + \mathcal{\bf D_{1}} \cdot {\bf \hat{p}_{i}}) \times (1 + \mathcal{\bf D_{2}} \cdot {\bf \hat{p}_{i}})$.
    We did this for all pixels over the sky and show the resultant map in Fig.~\ref{fig:signal-MultDipHyb-low}.
    Compare this map with the smoothed versions shows in the bottom row of Fig.~\ref{fig:all-maps}, noting the dashed lines which indicate the $30^*$ mask we applied.
    With this in mind, the high value of the Bayes factors likely arise because of the over-density at $(l,b) \approx (0^\circ, 30^\circ)$ and the under-densities along the galactic plane from about $l = 120^\circ$ to $l = 240^\circ$; this cannot be adequately captured by for instance the dipole $M_1$, but is better captured by the double dipole $M_2$.
    This applies both for Quaia low and Quaia high.
    Considering the coverage of the $|b| < 30^\circ$ mask, it is probable that this angle and those below it are not able to sufficiently remove the region of over-density.
    
\begin{figure}
        \centering
        \includegraphics[width=\linewidth]{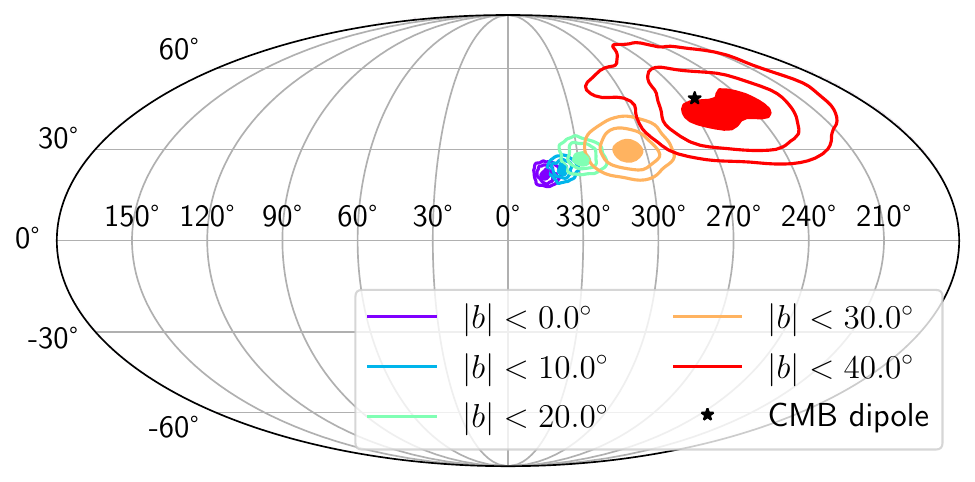}\vspace{2mm}
        {\renewcommand{\arraystretch}{1.5}
            \begin{tabular}{l S[table-format=2.0, round-precision=0] S[table-format=2.0, round-precision=0] S[table-format=2.0, round-precision=0] S[table-format=2.0,          round-precision=0] S[table-format=2.0, round-precision=0] S[table-format=1.0, round-precision=0]}
                \hline
                Amplitude $D$ ($\times10^3$) & {\color{ProjPurple}$37\substack{+4 \\ -4}$} & {\color{ProjBlue}$34\substack{+4 \\ -4}$} & {\color{ProjGreenDarker}$25\substack{+5 \\ -5}$} & {\color{ProjOrange}$17\substack{+6 \\ -6}$} & {\color{ProjRed}$11\substack{+6 \\ -5}$} & 8.0 \\[0.5mm]\hline
            \end{tabular}\vspace{4.3mm}
        }
        \includegraphics[width=\linewidth]{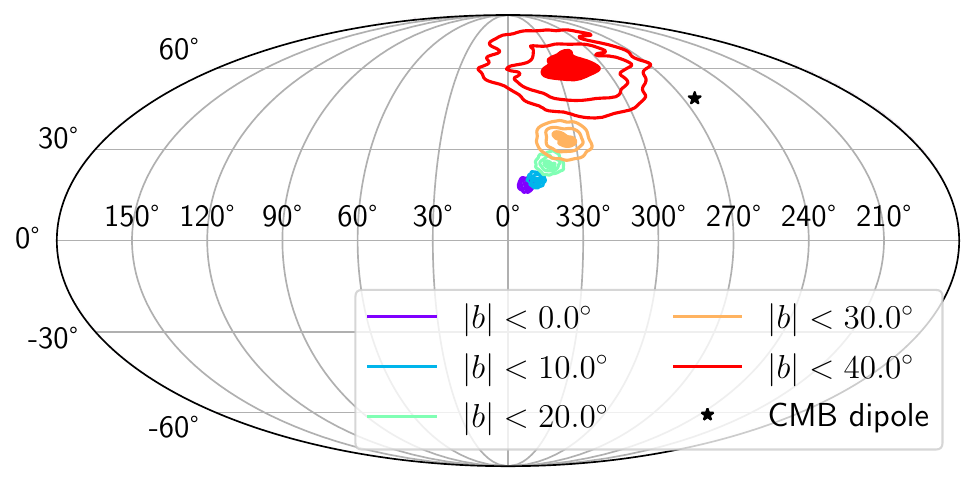}\vspace{2mm}
        {\renewcommand{\arraystretch}{1.5}
            \begin{tabular}{l S[table-format=2.0, round-precision=0] S[table-format=2.0, round-precision=0] S[table-format=2.0, round-precision=0] S[table-format=2.0,          round-precision=0] S[table-format=2.0, round-precision=0] S[table-format=1.0, round-precision=0]}
                \hline
                Amplitude $D$ ($\times10^3$) & {\color{ProjPurple}$48\substack{+4 \\ -3}$} & {\color{ProjBlue}$43\substack{+3 \\ -3}$} & {\color{ProjGreenDarker}$32\substack{+4 \\ -4}$} & {\color{ProjOrange}$20\substack{+4 \\ -4}$} & {\color{ProjRed}$12\substack{+4 \\ -3}$} & 6.8 \\[0.5mm]\hline
            \end{tabular}
        }\vspace{2mm}
        \caption{Projection of the posterior distribution for recovered dipole direction (using the Poissonian likelihood) onto a Mollweide projection. The graticules are in galactic coordinates, and the contours give intervals of 11.8\%, 39.4\% and 67.5\% of posterior mass, equivalent to $0.5\sigma$, $1\sigma$ and $1.5\sigma$ for a 2D Gaussian. The different colours correspond to different galactic plane cuts applied, as indicated by the legend. Amplitudes of the recovered dipole ($\times 10^3$) are also tabulated beneath the plots with the same colour code as the sky projections. The listed uncertainties give an interval containing 95\% of the posterior mass, equivalent to $2\sigma$ for a 1D Gaussian. The key point is that as increasingly conservative masks are used, the recovered and CMB dipoles more closely align for Quaia low. \textit{Top:} Quaia low. \textit{Bottom:} Quaia high.}
        \label{fig:mask-variance}
    \end{figure}

    We therefore are of the view that systematic effects arising from the Quaia selection function impact our analysis.
    These systematic effects -- whether arising from inadequate consideration of dust extinction effects, stellar contaminants, etc. -- likely contribute to over-dense regions near the galactic centre, which give spurious fits as far as probing the distribution of distant quasars is concerned.
    We can better understand the effect of this dubious over-density by examining how the direction and amplitude of the recovered dipole of model $M_1$ changes with galactic mask angle.
    This is illustrated in Fig.~\ref{fig:mask-variance}.
    Looking at the top pane (Quaia low), as the region of over-density near the galactic centre is progressively masked out with increasing galactic latitude $b$, the recovered dipole direction shifts towards the CMB dipole.
    For a $40^\circ$ galactic plane mask (red contours), the recovered direction is consistent with the CMB direction within $\approx 0.5 \sigma$, although we note that at this stage the uncertainty in the fitted parameters has drastically increased because more than half of the sky has been masked.
    With respect to the dipole amplitudes, these are shown for each mask in the single row below the Quaia low sky projection of Fig.~\ref{fig:mask-variance}.
    As the over-dense region is masked out, the recovered dipole amplitude approaches the CMB dipole amplitude.
    Interestingly, for a $40^\circ$ galactic plane mask, the recovered dipole amplitude $D \times 10^3 \approx 11\substack{+6 \\ -5}$ is consistent with the CMB dipole amplitude $D \times 10^3 \approx 8$ given the $2\sigma$ uncertainties, which sheds light on why $M_6$ is the preferred model in this regime.
    We interpret this as signifying that the spurious over-densities are filtered out with the $40^\circ$ mask, and so the prevailing model is simply a dipole consistent with the CMB dipole.

    We also attempted to remove this over-dense region -- while maximising the final number of sources analysed -- by imposing a  $4\text{ sr}$ circular mask centered at $(l, b)=(0^\circ,0^\circ)$ on top of the $30^\circ$ mask.
    This mask has been labelled as $30^*$ in Tables~\ref{tab:bayes-low-p2p}--\ref{tab:bayes-high-poisson}. 
    Note that this mask covers regions beyond the $|b| < 40^\circ$ limit and hence in principle is better in removing the over-densities concentrated in the northern hemisphere along the line of $l \approx 0^\circ$.
    Still, for Quaia low, the Bayes factor for the kinematic dipole ($M_6$) is about equal to that of the double dipole ($M_2$) where the $30^*$ mask is employed, suggesting they have equivalent explanatory power.\
    This is likely because the $30^*$ mask is being influenced by other over-densities and under-densities, which are better filtered out by the $40^\circ$ mask.
    We therefore find that the $40^\circ$ mask is most apt for genuinely interpreting the cosmic dipole in Quaia low.
    
    Quaia high cannot be viewed with the same interpretation.
    Looking at the bottom pane of Fig.~\ref{fig:mask-variance}, the recovered dipole for Quaia high drifts away from the CMB dipole and towards $(l,b) \approx (330^\circ, 60^\circ)$ with high galactic masks, which it reaches by $|b| < 40^\circ$.
    Moreover, the CMB and inferred dipole amplitudes do not agree with each other within $2\sigma$, even with the $40^\circ$ mask.
    Looking at the Bayes factors, the favoured hypothesis with a $40^\circ$ mask is the kinematic direction ($M_4$), but only marginally; the next-best model, $M_1$, is only 0.2 to 0.3 log units behind.
    This is only a superficial, `bare mention' of support for $M_4$ over $M_1$ \citep{kass1995}.
    We therefore form the view that, while a dipole is being inferred in the sample, its parameters cannot be constrained.
    That is, the data is insufficiently clear on whether this dipole aligns with the CMB dipole, or has a preferred direction somewhere away from the CMB dipole.
    This is likely happening because, while $M_4$ has one less parameter than $M_1$ and hence a more favourable Occam factor, it also has less explanatory power.
    On net, the models balance out in terms of support.

    We are hesitant to draw genuine conclusions from Quaia high because of the building evidence that the sample suffers more severely from contamination than Quaia low.
    Inspection of Fig.~\ref{fig:mask-variance} lends support to this, but as an additional check, we tested one final model on both samples.
    Suppose that Quaia high is in fact contaminated, such that the sample consists of some genuine dipole component aligned with that of Quaia low (L), as well as a contaminated component.
    If so, then using our Poissonian methodology, the rate parameter describing Quaia high (H) for the $i$-th pixel would be
    \begin{align}
        (\lambda_i)_{\text{H}} &= (\lambda_i)_{\text{L}} + (\lambda_i)_{\text{H} - \text{L}} \label{eq:fitboth1} \\
                               &= \bar{N}_\text{L} (1 + \mathbf{D}_{\text{L}} \cdot \mathbf{\hat{p}_i}) + \bar{N}_{\text{H} - \text{L}} (1 + \mathbf{D}_\text{C} \cdot \mathbf{\hat{p}_i}) \label{eq:fitboth2}
    \end{align}
    where $\text{H} - \text{L}$ refers to the `high minus low' sample and C refers to the contaminated dipole component.
    The `high minus low' sample is simply a catalogue we constructed containing the sources in Quaia high that are not in Quaia low.
    In this model, we fit both catalogues simultaneously: we use the rate parameter $(\lambda_i)_\text{L}$ to fit a dipole to Quaia low, which is our genuine component, and use the rate parameter $(\lambda_i)_{\text{H}}$ to fit the Quaia high sample.
    Accordingly, the total likelihood is the sum of the two likelihoods for both catalogues, i.e.
    \begin{equation}
        \ln \mathcal{L}_{\text{tot.}} = \sum_{i=1}^{n_{\text{pix.}}} \ln P(N_i | \, \mathbf{\Theta}_{\text{L}}) +  \sum_{i=1}^{n_{\text{pix.}}} \ln P(N_i | \, \mathbf{\Theta}_{\text{H}}). \label{eq:fitboth3}
    \end{equation}
    In this case, the set of parameters $\mathbf{\Theta}_{\text{L}} = \{ \bar{N}_{\text{L}}, D_{L}, l_{L}, b_{L} \}$ and $\mathbf{\Theta}_{\text{H}} = \{ \mathbf{\Theta}_{\text{L}}, \bar{N}_{\text{H} - \text{L}}, D_\text{C}, l_{\text{C}}, b_{\text{C}} \}$, since the rate parameter for Quaia high now depends on the parameters of the dipole for Quaia low and the parameters for the contaminated `high minus low' component.

\begin{figure}
        \centering
        \includegraphics[width=\linewidth]{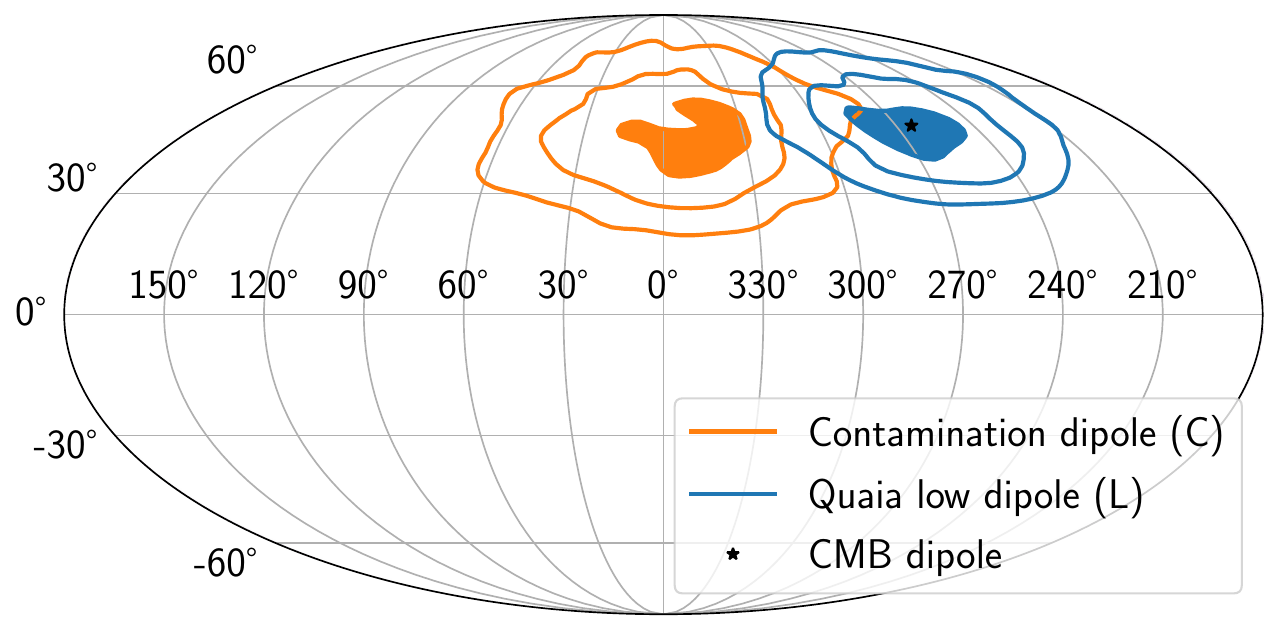}
        \caption{Projection of the 2D posterior distribution for the Quaia low dipole and the contaminated dipole, as defined in equations~\eqref{eq:fitboth1}--\eqref{eq:fitboth3}, using a $40^\circ$ mask. The contours give intervals of 11.8\%, 39.4\% and 67.5\% of posterior mass, equivalent to $0.5\sigma$, $1\sigma$ and $1.5\sigma$ for a 2D Gaussian.}
        \label{fig:cont-dipole}
    \end{figure}

    Testing this on a $40^\circ$ mask yielded the posterior distributions seen in Fig~\ref{fig:cont-dipole}.
    Evidently, the Quaia low dipole is recovered well (cf. Fig.~\ref{fig:mask-variance}), and the bulk of the posterior mass coincides with the CMB dipole.
    In contrast, the contamination dipole, which to reiterate has a rate parameter described by $\bar{N}_{\text{H} - \text{L}} (1 + \mathbf{D}_\text{C} \cdot \mathbf{\hat{p}_i})$, coincides with the region of over-density near the galactic centre.
    This is instructive insofar that it lends further evidence to their being sources near the galactic centre which significantly contaminate Quaia high.

    
    To summarise all the above, we form the view that while the over-densities are removed with the $|b| < 40^\circ$ mask for Quaia low, they are present beyond this limit for Quaia high.
    Thus, we use the $40^\circ$ mask to infer the cosmic dipole in Quaia low, and more broadly the Quaia sample.
    This is predicated on its ability to remove most of the clustering issues arising from the selection function.
    For Quaia high, we can at best infer that there is a dipole, but not its parameters.
    There is strong reason to believe that this is because the sample is contaminated by sources not part of the quasar background.

    \subsection{Dipole estimation from the results}
        Having interpreted our results, we present our final determination of the cosmic dipole in Quaia.
        To reiterate the foregoing, we take the result from the $40^\circ$ mask from Quaia low
        and withhold any conclusions regarding Quaia high.
    
        For Quaia low, a dipole aligned both in direction and magnitude with the CMB dipole ($M_6$) best accounts for the results.
        This model has positive support over the next-favoured model $M_5$ ($\ln B_{65} = 2.6$), which only fixes the dipole magnitude to that expected from the CMB.
        Accordingly, the inferred distribution of quasars in Quaia low is consistent with the kinematic interpretation of the CMB, and thus is consistent with the cosmological principle.
        

    \subsection{Impact of priors on results}
        It is worth noting that because the marginal likelihood in Bayes's theorem is an integral over all parameter space, it is generally sensitive to the choice of prior.
        This could have an effect on the inferences made when evaluating competing hypotheses.
        Thus, we also investigated whether or not our conclusions still hold with a narrower prior on the dipole amplitude.
        Specifically, we sampled the dipole amplitude according to $D \sim [0,0.1]$ and the quadrupole amplitude according to $\tilde{D} \sim [-0.1, 0]$.
        These ranges of values are one order of magnitude smaller than our previous choices.

        With more restrictive priors, we find that our conclusions do not change; that is, our findings are not strongly sensitive to the choice of prior.
        In general, the marginal likelihood for the dipole hypothesis ($M_1$), quadrupole hypothesis ($M_3$) and kinematic direction hypothesis ($M_4$) increased with respect to the other hypotheses.
        With a $|b| < 40^\circ$ galactic plane mask on Quaia low, the Bayes factor for $M_1$ moved from $2.8 \to 5.2$, $0.5 \to 2.9$ for $M_3$ and $5.8 \to 8.0$ for $M_4$.
        This is with respect to the point-by-point approach, but the trend of increasing Bayes factors was similar for the Poissonian approach.
        While $M_4$ is slightly more favoured than it was hitherto, $M_6$ is still the dominant hypothesis (recall it has a Bayes factor of 10, as in Table~\ref{tab:bayes-low-p2p}).
        Similar changes occur for Quaia high.
        $M_4$ remains the favoured hypothesis, but with only marginally more support than $M_1$.
        However, since the marginal likelihood for $M_4$ increases while $M_5$ and $M_6$ are fixed between the two prior functions, $M_4$ prevails more significantly over the other kinematic hypotheses.
        None the less, we reserve drawing further conclusions from Quaia high for the reasons mentioned in Section~\ref{sub:prevailing_double_dipole}.
        Thus, even with more restrictive priors, the best explanation of the data is a dipole consistent with the CMB dipole.
    
    \subsection{Final thoughts and future outlook}
        In this work, we presented a Bayesian analysis of the distribution of quasars in the Quaia catalogue of \citet{storeyfisher2023quaia}.
        Analysing both the Quaia low and Quaia high samples, there is substantial evidence that systematic effects arising from the construction of the selection function introduce spurious over-densities and under-densities into the catalogue, contaminating it.
        This affects the inference of the cosmic dipole for both samples.
        After masking the contaminated regions, the inferred dipole is consistent with the CMB dipole in both magnitude and direction for Quaia low, but the data affords insufficient clarity to define the parameters of the dipole for Quaia high.
        There, the dipole drifts along $l = 0^\circ$ towards high values of $b$ as more conservative galactic plane masks are applied, which we have identified as an effect of a large over-density in the northern hemisphere.
        However, taking the results together, the inferred dipole in the distribution of quasars is in agreement with the CMB dipole, and hence the cosmological principle.

        There are numerous avenues that can be pursued in the future to build on this result. These include:
        \begin{enumerate}
            \item \emph{Studying the redshift-based evolution of the dipole.}
            In this work, we examined the cosmic dipole of both Quaia samples without reference to the quasar redshift distribution.
            A redshift-binned selection function map for Quaia high has been released, in which a cut at $z=1.47$ was imposed to divide sources into those at a redshift lower and greater than this threshold \citep{alonso2023}.
            Future work might generate a number of source redshift bins, regenerating the selection map for each bin using the code released with \citet{storeyfisher2023quaia}, and then study the evolution of the cosmic dipole as a function of redshift. 
            This approach has been utilised by \cite{horstmann2022} in their study of the dipole evolution of Type Ia SNe.
            \item \emph{A joint analysis with other data-sets.}
            A joint analysis of Quaia with other all-sky catalogues such as radio galaxies, SNe and other quasar samples can give interesting insights into the overall matter dipole of the universe.
            It might also shed light on how systematic effects are influencing the recovered dipole for different catalogues.
            Such cross-sample studies have been performed by many different research groups \citep[see e.g.][]{secrest2022, darling2022, wagenveld2023}.
            \item \emph{A revisit of the Quaia selection function.} In this work, substantial masking of the galactic plane was employed to screen out over-densities and under-densities, which are suspected to arise not because of the Earth's peculiar motion, but because of systematic errors introduced by the selection function.
            Our primary concern is how the selection function appears to over-estimate source density near the galactic centre.
            These density fluctuations have limited our ability to determine the cosmic dipole of Quaia, chiefly by reducing the final number of sources which are analysed after masking.
            Hence a full appraisal of the selection function will be essential in determining the robustness of the cosmic dipole within the Quaia sample and its impact on the cosmological principle.
        \end{enumerate}


\section*{Acknowledgements}
    We first wish to extend our sincere gratitude to the authors of \citet{storeyfisher2023quaia} -- we thank them for making the Quaia catalogue publicly available and for their positive responses to our queries.
    We also wish to acknowledge and thank the insightful comments made by the anonymous referee.
    VM acknowledges Prof. Kulinder Pal Singh, IISER Mohali for his support; DST, Government of India for the INSPIRE-SHE scholarship and University of Sydney's internship program which made this work possible.
    OTO is supported by the Australian Government Research Training Program (RTP) Scholarship.
    This work made use of the Python packages \textsc{dynesty} \citep{skilling2004, skilling2006, dynesty-v2.1.2}, \textsc{healpy} \citep{Gorski2005,Zonca2019}, \textsc{numpy} \citep{harris2020}, \textsc{matplotlib} \citep{hunter2007}, \textsc{pandas} \citep{mckinney2010,reback2020}, \textsc{scipy} \citep{scipy2020} and \textsc{astropy} \citep{astropy2022}.

\section*{Data Availability}

The data used in this study will be made available with a reasonable request to the authors.



\bibliographystyle{mnras}
\bibliography{Quaia} 




\appendix

\section{Bayes Factors for Tested Hypotheses}
\label{sec:appendix}
Tables of Bayes factors for all the tested hypotheses are given on the following page. 
\onecolumn
\begin{table}
  \centering
   \begin{tabular}
   {l  S[table-format=7.1, round-precision=1] S[table-format=7.1,round-precision=1] S[table-format=7.1, round-precision=1] S[table-format=7.1,round-precision=1] S[table-format=7.1, round-precision=1] S[table-format=7.1, round-precision=1] }
\hline
\multirow{2}*{Hypothesis} & \multicolumn{6}{c}{Galactic mask angle $b^\circ$}\\
  & 0 & 10 &20&30&40&30*\\
  \hline
$M_{0}$ (Null) & 0.000000 & 0.000000 & 0.000000 & 0.000000 & 0.000000 & 0.000000 \\
$M_{1}$ (Dipole) & 130.600000 & 109.500000 & 49.400000 & 14.400000 & 2.800000 & 11.307116 \\
$M_{2}$ (Double Dipole) & 175.003049\cellcolor{black!10} & 130.598289\cellcolor{black!10} & 57.706860\cellcolor{black!10} & 18.877400\cellcolor{black!10} & 0.229165 & 14.474102 \\
$M_{3}$ (Quadrupole) & 44.300000 & 21.100000 & 10.200000 & 6.200000 & 0.500000 & 11.974102 \\
$M_{4}$ (Kinematic Direction) & 23.418572 & 30.460794 & 24.045107 & 12.776030 & 5.750978 & 13.213199 \\
$M_{5}$ (Kinematic Velocity) & 50.369723 & 45.749313 & 28.687372 & 13.996625 & 7.429802 & 11.769300 \\
$M_{6}$ (Kinematic Dipole) & 23.352709 & 26.983822 & 22.901712 & 15.494468 & 10.001544\cellcolor{black!10} & 14.638968\cellcolor{black!10} \\
\hline
 \end{tabular}
 \caption{Table of Bayes Factors for different hypotheses and galactic masks using the Quaia low catalogue with the point-by-point analysis. Here, 30$^*$ represents the combination of a $30^\circ$ mask and a $4\,\text{sr}$ circular mask centered at the $(l^\circ, b^\circ)=(0,0)$. The highlighted cell represents the model with the highest Bayes factor, indicating it has the strongest level of support.
 }
 \label{tab:bayes-low-p2p}
 \end{table}

 \begin{table}
  \centering
   \begin{tabular}
   {l  S[table-format=7.1, round-precision=1] S[table-format=7.1,round-precision=1] S[table-format=7.1, round-precision=1] S[table-format=7.1,round-precision=1] S[table-format=7.1, round-precision=1] S[table-format=7.1, round-precision=1] }
\hline
\multirow{2}*{Hypothesis} & \multicolumn{6}{c}{Galactic mask angle $b^\circ$}\\
  & 0 & 10 &20&30&40&30*\\
  \hline
$M_{0}$ (Null) & 0.000000 & 0.000000 & 0.000000 & 0.000000 & 0.000000 & 0.000000 \\
$M_{1}$ (Dipole) & 131.400000 & 109.900000 & 49.900000 & 14.500000 & 3.500000 & 11.700000 \\
$M_{2}$ (Double Dipole) & 175.613985\cellcolor{black!10} & 132.002973\cellcolor{black!10} & 58.892644\cellcolor{black!10} & 19.991748\cellcolor{black!10} & 1.447581 & 15.200000\cellcolor{black!10} \\
$M_{3}$ (Quadrupole) & 45.500000 & 22.100000 & 11.200000 & 7.400000 & 1.600000 & 13.100000 \\
$M_{4}$ (Kinematic Direction) & 23.783909 & 30.697844 & 24.453275 & 12.941546 & 6.114179 & 13.426083 \\
$M_{5}$ (Kinematic Velocity) & 50.924243 & 46.379544 & 29.469394 & 14.076147 & 7.909093 & 12.071233 \\
$M_{6}$ (Kinematic Dipole) & 23.725727 & 27.498682 & 23.584607 & 15.966434 & 10.524382\cellcolor{black!10} & 15.197638\cellcolor{black!10} \\
\hline
 \end{tabular}
 \caption{As for Table~\ref{tab:bayes-low-p2p} but with the Poisson statistics.
 }
 \label{tab:bayes-low-poisson}
 \end{table}

 \begin{table}
  \centering
   \begin{tabular}
   {l  S[table-format=7.1, round-precision=1] S[table-format=7.1,round-precision=1] S[table-format=7.1, round-precision=1] S[table-format=7.1,round-precision=1] S[table-format=7.1, round-precision=1] S[table-format=7.1, round-precision=1] }
\hline
\multirow{2}*{Hypothesis} & \multicolumn{6}{c}{Galactic mask angle $b^\circ$}\\
  & 0 & 10 &20&30&40&30*\\
  \hline
$M_{0}$ (Null) & 0.000000 & 0.000000 & 0.000000 & 0.000000 & 0.000000 & 0.000000 \\
$M_{1}$ (Dipole) & 375.900000 & 308.500000 & 146.400000 & 49.600000\cellcolor{black!10} & 21.000000 & 18.618201 \cellcolor{black!10} \\
$M_{2}$ (Double Dipole) & 426.827565\cellcolor{black!10} & 329.175231\cellcolor{black!10} & 156.449248\cellcolor{black!10} & 49.618615\cellcolor{black!10} & 20.283198 & 16.323110 \\
$M_{3}$ (Quadrupole) & 64.400000 & 27.500000 & 10.200000 & 1.600000 & 0.000000 & 6.818201 \\
$M_{4}$ (Kinematic Direction) & 32.569929 & 43.173853 & 43.572661 & 29.150867 & 21.325630 \cellcolor{black!10} & 16.010938 \\
$M_{5}$ (Kinematic Velocity) & 98.654164 & 88.895461 & 57.955084 & 31.386007 & 21.076589 & 16.837082 \\
$M_{6}$ (Kinematic Dipole) & 30.336517 & 35.703677 & 34.850180 & 26.080769 & 20.397992 & 17.443674 \\
\hline
 \end{tabular}
 \caption{As for Table~\ref{tab:bayes-low-p2p} but with Quaia high.}
 \label{tab:bayes-high-p2p}
 \end{table}

 \begin{table}
  \centering
   \begin{tabular}
   {l  S[table-format=7.1, round-precision=1] S[table-format=7.1,round-precision=1] S[table-format=7.1, round-precision=1] S[table-format=7.1,round-precision=1] S[table-format=7.1, round-precision=1] S[table-format=7.1, round-precision=1] }
\hline
\multirow{2}*{Hypothesis} & \multicolumn{6}{c}{Galactic mask angle $b^\circ$}\\
  & 0 & 10 &20&30&40&30*\\
  \hline
$M_{0}$ (Null) & 0.000000 & 0.000000 & 0.000000 & 0.000000 & 0.000000 & 0.000000 \\
$M_{1}$ (Dipole) & 376.000000 & 308.100000 & 146.500000 & 49.300000 & 21.200000 & 18.778620 \cellcolor{black!10} \\
$M_{2}$ (Double Dipole) & 426.791312\cellcolor{black!10} & 328.539441\cellcolor{black!10} & 155.837352\cellcolor{black!10} & 50.917748\cellcolor{black!10} & 20.201345 & 16.176393 \\
$M_{3}$ (Quadrupole) & 64.400000 & 27.700000 & 10.200000 & 2.300000 & 0.200000 & 7.178620 \\
$M_{4}$ (Kinematic Direction) & 32.519147 & 43.213933 & 43.596229 & 29.243384 & 21.417602 \cellcolor{black!10} & 16.346723 \\
$M_{5}$ (Kinematic Velocity) & 98.306560 & 88.593329 & 57.663709 & 31.606552 & 21.013921 & 16.615128 \\
$M_{6}$ (Kinematic Dipole) & 30.195953 & 35.724368 & 34.653692 & 26.172292 & 20.359357 & 17.371709 \\
\hline
 \end{tabular}
  \caption{As for Table~\ref{tab:bayes-low-p2p} but with Quaia high and the Poisson statistics.}
  \label{tab:bayes-high-poisson}
 \end{table}
 


\bsp	
\label{lastpage}
\end{document}